\documentclass[fleqn,usenatbib,useAMS]{mnras}

\usepackage[T1]{fontenc}
\usepackage{ae,aecompl}
\usepackage{newtxtext,newtxmath}

\usepackage{url}
\usepackage{lscape}
\usepackage{verbatim}
\usepackage{amsmath}
\usepackage{graphicx}
\usepackage{multicol}       
\usepackage{pdflscape}
\usepackage{tablefootnote}

\def\beq{\begin{equation}}
\def\eeq{\end{equation}}
\def\alwaysmath#1{{\ifmmode{#1}\else{$#1$}\fi}}
\def\iso#1#2{\mbox{${}^{#2}{\rm #1}$}}
\def\he#1{\iso{He}{#1}}

\def\10830{{He~I $\lambda$10830}}
\def\3889{{He~I $\lambda$3889}}

\providecommand{\HA}{H$\alpha$}                                 
\providecommand{\HB}{H$\beta$}                                  
\providecommand{\HG}{H$\gamma$}                                 
\providecommand{\HD}{H$\delta$}                                 

\newcommand\hi{H\,{\sc I}}

\title
[A comprehensive chemical abundance analysis of AGC~198691]{
A comprehensive chemical abundance analysis of the extremely metal poor Leoncino Dwarf galaxy (AGC~198691)
}

\author[E.~Aver, D.A.~Berg, A.S.~Hirschauer, K.A.~Olive, R.W.~Pogge, N.S.J.~Rogers, J.J.~Salzer, \& E.D.~Skillman]{
Erik~Aver$^{1}$,
Danielle~A.~Berg$^{2}$,
Alec~S.~Hirschauer$^{3}$,
Keith~A.~Olive$^{4,5,6}$, 
Richard~W.~Pogge$^{7,8}$,
\newauthor
Noah~S.~J.~Rogers$^{4,6}$,
John~J.~Salzer$^{9}$,
Evan~D.~Skillman$^{4,6}$
\thanks{
e-mail addresses: 
\begin{flushleft}
\href{mailto:aver@gonzaga.edu}{aver@gonzaga.edu}, \href{mailto:daberg@austin.utexas.edu}{daberg@austin.utexas.edu},
\href{mailto:ahirschauer@stsci.edu}{ahirschauer@stsci.edu},
\href{mailto:olive@umn.edu}{olive@umn.edu},
\href{mailto:pogge.1@osu.edu}{pogge.1@osu.edu},
\href{mailto:roge0291@umn.edu}{roge0291@umn.edu},
\href{mailto:josalzer@indiana.edu}{josalzer@indiana.edu},
\href{mailto:skill001@umn.edu}{skill001@umn.edu}
\end{flushleft}
}
\\
$^{1}$Department of Physics, Gonzaga University, 
502 E Boone Ave., Spokane, WA 99258 \\
$^{2}$Astronomy Department, University of Texas at Austin,
 Austin, TX 78712 \\
$^{3}$Space Telescope Science Institute, 3700 San Martin Drive, 
 Baltimore, MD 21218 \\
$^{4}$School of Physics and Astronomy, University of Minnesota, 
116 Church St. SE, Minneapolis, MN 55455 \\
$^{5}$William I. Fine Theoretical Physics Institute, University of Minnesota, 
116 Church St. SE, Minneapolis, MN 55455 \\
$^{6}$Minnesota Institute for Astrophysics, University of Minnesota, 
116 Church St. SE, Minneapolis, MN 55455 \\
$^{7}$Department of Astronomy, The Ohio State University, 
140 W 18th Ave., Columbus, OH 43210 \\
$^{8}$Center for Cosmology \& AstroParticle Physics, The Ohio State University, 
191 West Woodruff Ave., Columbus, OH 43210 \\
$^{9}$Department of Astronomy, Indiana University, 
727 East Third St., Bloomington, IN 47405 \\
}

\begin{document}
\label{firstpage}
\pagerange{\pageref{firstpage}--\pageref{lastpage}}
\maketitle

\begin{abstract}
We re-examine the extremely metal-poor (XMP) dwarf galaxy AGC~198691 using a high quality spectrum obtained by the LBT's MODS instrument.  Previous spectral observations obtained from KOSMOS on the Mayall 4-m and the Blue Channel spectrograph on the MMT 6.5-m telescope did not allow for the determination of sulfur, argon, or helium abundances.  We report an updated and full chemical abundance analysis for AGC~198691, including confirmation of the extremely low ``direct" oxygen abundance with a value of 12 + log(O/H) = 7.06 $\pm$ 0.03.  AGC~198691's low metallicity potentially makes it a high value target for helping determine the primordial helium abundance ($Y_p$).  Though complicated by a Na~I night sky line partially overlaying the He~I $\lambda$5876 emission line, the LBT/MODS spectrum proved adequate for determining AGC~198691's helium abundance.  
We employ the recently expanded and improved model of \citet{AOS5}, incorporating higher Balmer and Paschen lines, augmented by the observation of the infrared helium emission line \10830 obtained by \citet{hcpb}.  
Applying our full model produced a reliable helium abundance determination, consistent with the expectation for its metallicity.
Although this is the lowest metallicity object with a detailed helium abundance, unfortunately, due to its faintness (EW(H$\beta$) $<$ 100 \AA ) and the compromised He~I $\lambda$5876, the resultant uncertainty on the helium abundance is too large to allow a significant improvement on the measurement of $Y_p$. 
\end{abstract}




\section{Introduction}\label{intro}

\subsection{Extremely Metal Poor Galaxies}

Extremely metal-poor galaxies (XMPs) have been defined as galaxies with 
metal abundances roughly 5\% of the solar value or less, 12 + log(O/H) $\le$ 7.35
\citep{guseva2015,mcquinn2020}.  This definition has been adopted to characterize the most extreme systems amongst the growing number of metal poor galaxies known \citep{mcquinn2020}.  
These galaxies are of importance for studying the 
star formation process at low metallicity, which is relevant for understanding
star formation in the early universe. Furthermore, when an XMP galaxy hosts a
high surface brightness \ion{H}{ii} region, then these galaxies can be
used to determine the primordial helium abundance ($Y_p$).

The number of actively star forming XMP galaxies remains
small due to various limitations in our ability to identify them
\citep[see][]{sanchez2017}.
Following the analysis presented in \citet{mcquinn2020}, to date, there appear to be two different formation channels for XMP
galaxies.  The first is simply the result of the relationship between 
a galaxy's stellar-mass and its metallicity \citep[e.g.,][]{tremonti2004, berg2012}.  This relationship implies that the most 
metal-poor galaxies are the least luminous, and therefore are difficult
to identify in galaxy surveys.  A second channel for XMP formation has 
been identified with the recent infall of relatively pristine gas onto 
a low mass galaxy \citep[e.g.,][]{ekta2010}.  This infall results in a depression of the gas 
metallicity and this can drive a galaxy into the regime of XMP galaxies.

There has been recent progress in identifying XMP galaxies in both of
these categories.  For example, 
\citet{skillman2013, hirschauer2016, hsyu2018, james2017, yang2017, senchyna2019, senchyna2019b, kojima2020, pustilnik2020} have all discovered XMP
galaxies by obtaining spectra of inherently low luminosity galaxies.
\cite{itg12, guseva2017, izotov2018} have identified very high surface brightness
star forming regions in XMP galaxies that tend to be preferentially 
in the recent infall category.

The lower limit on the metallicity of a star forming region is also
of interest in constraining our knowledge of the evolution of galaxies.
The current record holder has been identified by 
\citet{izotov2018} to have
an oxygen abundance of 12 + log (O/H) $=$ 6.98, which corresponds to
roughly 2\% of the solar value \citep[adopting a value of 8.69][]{asplund2009}.

\subsection{The Primordial Helium Abundance}

Among the element abundances observed in XMP galaxies, \he4 offers a window into the physics of the early universe, as these abundances can be used to determine the primordial helium abundance and test standard big bang nucleosynthesis (SBBN) \citep{wssok,osw,pis,cfoy,coc18,fields2020}. 
With the cosmic microwave background (CMB) determination of the baryon density \citep{planck18}, SBBN 
is effectively a parameter-free theory \citep{cfo2}. The primary uncertainty in the SBBN calculation of the \he4 abundance is the neutron mean life,
 $\tau_n = 879.4 \pm 0.6$ s \citep{rpp}, and the predicted primordial abundance is $Y_p = 0.2469 \pm 0.0002$ \citep{fields2020,Yeh:2020mgl}, i.e., with less than 0.1\% accuracy. 
 In contrast, the accuracy of the predicted deuterium abundance is about 4\%, D/H $= (2.51 \pm 0.11) \times 10^{-5}$ \citep{Yeh:2020mgl}.  
If precise abundance measurements are available, \he4 and SBBN can be used to probe
the physics beyond the standard model \citep{cfos}, for example, new particle degrees of freedom, often scaled as the number of light neutrinos, $N_\nu$. 

Currently, the primordial helium abundance, $Y_p$ is best determined by fitting the helium abundances of individual objects, ideally XMP galaxies, versus a measurement of metallicity (e.g., oxygen), and extrapolating back to zero metallicity \citep{ptp74}. 
\he4 abundance determinations come from measurements of emission line spectra of extragalactic H~{\scriptsize II} regions. 
However, obtaining an abundance from these data requires careful modeling, and there are significant sources of potential systematic uncertainty \citep{os01,os04,plp07,its07,AOS}. 
 The use of Monte Carlo methods helps evaluate the fit between the data and theoretical model and captures the uncertainties on the physical model parameters from that data lensed through the model \citep{AOS2}. 
\citet{AOS3} found that most of the available observations are statistically inconsistent with the models used to extract abundances and employed corresponding quality cuts on their datasets.  Recently, other analyses have found similar inconsistencies \citep{ftdt2,hcpb}.
Whether these incompatibilities are due to deficiencies in the model, or the data, or both, is an open question which can only be answered through obtaining new observations. 

\citet{AOS5} have addressed both the model and data in order to increase the accuracy of \he4 abundance determinations.  They improved the treatment of the collisional excitation of H~I and of underlying stellar absorption, as well as the approach for the blended line \3889 with H8, including its underlying absorption and radiative transfer.  In conjunction, they also expanded the model to utilize additional emission lines, not previously employed in helium abundance analyses.  

The Large Binocular Telescope (LBT)'s Multi-Object Double Spectrograph \citep[LBT/MODS,][]{pogge2010} allows for optical \& near-IR spectrum from the UV atmospheric cutoff to 1 $\mu$m.
Enabled by the LBT/MODS spectrograph's higher resolution, higher sensitivity, and broader wavelength coverage, \citet{AOS5} were able to employ previously unused weaker helium, higher Balmer, and NIR Paschen emission lines observed for the XMP galaxy Leo~P.  Based on the gains demonstrated in \citet{AOS4} from the \10830 emission line, the LBT Utility Camera in the Infrared \citep[LUCI,][]{seifert2003} was used to obtain a near-IR spectrum, from 0.95 to 1.35 $\mu$m, providing the high-impact \10830 emission line.  The combined effect of the model improvements, additional optical emission lines, and IR \10830 emission line was to reduce the uncertainty on the helium abundance determination of Leo~P by 70\% \citep{AOS5}, leading to an improvement in the uncertainty in $Y_p$ of 15\% (from $Y_p = 0.2449 \pm 0.0040$ to $0.2453 \pm 0.0034$)
simply from a reanalysis and addition of a single object.

This work takes a similar approach for the XMP galaxy AGC~198691
\citep{hirschauer2016}, leveraging a high-resolution LBT/MODS optical spectrum to incorporate higher Balmer and NIR Paschen emission lines, combined with a \10830 measurement from \citet{hcpb}, to determine the helium abundance for AGC~198691.  As was the case for Leo P \citep{AOS5}, the benefits of our improved methodology are significant.  The LBT/MODS spectrum allows for a confirmtation of the extremely low oxygen abundance determination, as well as the first determination of the helium abundance of AGC~198691.  

\bigskip

This paper is organized as follows.
First, in Section \ref{AGC}, we provide new observational data from AGC~198691. 
An analysis of the new data is made in section \ref{Results}.  In \S \ref{abundances}, the electron temperature and corresponding abundances are derived.  \he4 requires different methodology.  That methodology is presented and the helium abundance determined in \S \ref{He}.  
Finally, Section \ref{Conclusion} offers a discussion of the results and further prospects for improvement.

\section{New LBT Observations of AGC~198691} \label{AGC}

\subsection{Background}

AGC~198691 was discovered in the wide-field, unbiased \hi\ 21 cm ALFALFA survey \citep{giovanelli2005, haynes2011}.
AGC~198691 was originally recognized as being a galaxy of interest through its inclusion in the Survey for H~I in Extremely Low-mass Dwarfs (SHIELD) sample \citep{cannon2011}.  
Optical spectroscopic observations of the lone H~II region in AGC~198691 were first reported in \citet{hirschauer2016}.  Three spectral observations were obtained.  One using the KPNO Ohio State Multi-Object Spectrograph (KOSMOS) on the Mayall 4-m and two nights using the Blue Channel spectrograph on the MMT 6.5-m telescope.  
The oxygen abundance in AGC~198691 was measured to be 12 + log(O/H) = 7.02 $\pm$ 0.03 \citep{hirschauer2016}, based on composite spectrum from two night's observations using the MMT.  
At the time of publication, this was the lowest metallicity XMP known.  

\citet{hsyu2018} subsequently obtained a spectrum of AGC~198691 (referred to
as J0943+3326 in that work) with the Keck observatory LRIS instrument and derived
an oxygen abundance of 12 + log(O/H) $=$ 7.16 $\pm$ 0.07.  This confirmed
AGC~198691 as an XMP galaxy, but raised questions about the record holder status.
In the interim, \citet{itgl18} have discovered J0811+4730 with an oxygen
abundance of 12 + log(O/H) $=$ 6.98 $\pm$ 0.02.

AGC~198691 has recently been observed with the Hubble Space Telescope by \citet{mcquinn2020}.  They determine a distance of 12.1$^{+1.7}_{-3.4}$ Mpc
from the observation of the tip of the red giant branch with a resultant
estimate of stellar mass of 7.3 $\times$ 10$^5$ M$_\odot$.  This result 
implies that AGC~198691 is consistent with the stellar mass-metallicity
relationship established by nearby galaxies \citep{berg2012}, similar
to Leo~P \citep{skillman2013,mcquinn2015a}.  \citet{mcquinn2020} quoted 12 + log(O/H) $=$ 7.12 $\pm$ 0.04, based on a preliminary analysis of the observation reported in this work, but using an earlier version of the data reduction pipeline.  

\subsection{LBT Observations} \label{LBT}

After the MMT spectrum revealed the very low oxygen abundance of AGC~198691, we acquired an LBT observation in order to obtain a higher signal-to-noise spectrum and to detect
weak emission lines over a larger wavelength range.
A spectrum was obtained with the Multi-Object Double Spectrograph on
the Large Binocular Telescope \citep[LBT/MODS,][]{pogge2010} on 2015 December 6 under relatively clear
skies and approximately 1$\arcsec$ seeing.
The blue side of MODS1 covers a wavelength range of 3200 to 5650 \AA\ with a 400 l mm$^{-1}$ grating providing a resolution of 2.4 \AA.
The red side of MODS1 covers a wavelength range of 5650 to 10000 \AA\ with a 250 l mm$^{-1}$ grating providing a resolution of 3.4 \AA.
Bias frames, flat-field lamp images, and sky flats were obtained.
We observed GD71 as the standard star with a
5x60\arcsec\  spectrophotometric slit mask near the parallactic
angle.  The sensitivity function is determined using the HST CALSPEC flux tables \citep[e.g.,][]{bohlin10} which extend further into the UV and Near-IR than the original \citet{oke90} standard star fluxes.

\begin{figure*}
\centering
\includegraphics[width=\textwidth,trim=10mm 100mm 10mm 30mm,clip]{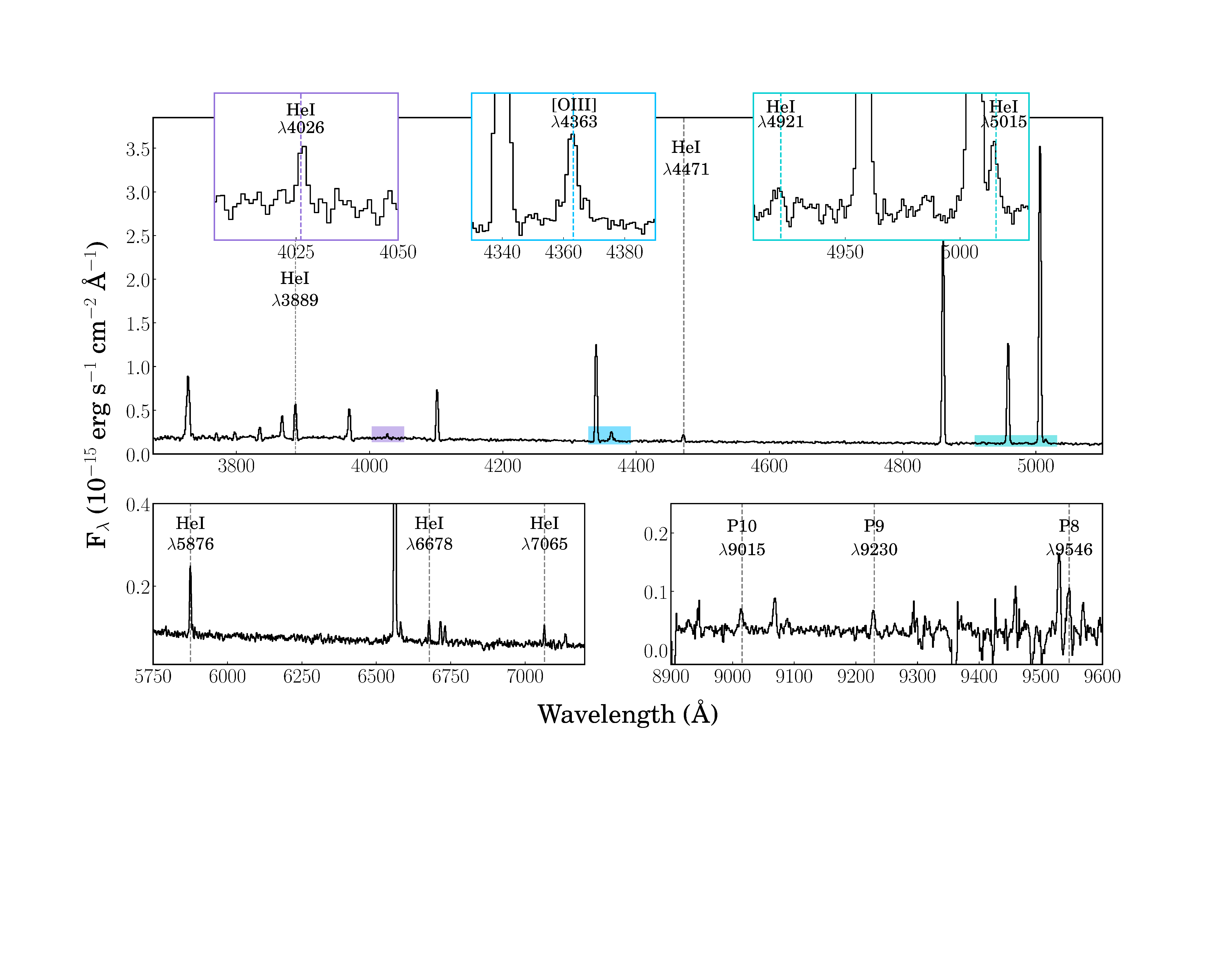}
\caption{
New LBT spectrum of AGC 198691 emphasizing weaker emission lines critical to the reported abundance analysis.  Specifically, \ion{O}{iii} $\lambda$4363 allowing a temperature measurement for a direct oxygen abundance and several \ion{He}{i} and \ion{H}{i} emission lines which are used to solve for the reddening, radiative transfer, physical conditions, and helium abundance simultaneously.
}
\label{Spectrum}
\end{figure*}

A total exposure time of 120 minutes was obtained with the MODS1 spectrograph
in six 20 minute exposures.
The extraction window was chosen to optimize the signal-to-noise in the faintest emission lines but broad enough to ensure a high-fidelity, robust measurement ($\sim$22 0.12$^{\prime\prime}$ pixels). 
The spectrum is shown in Figure~\ref{Spectrum}.

\subsection{Emission line fluxes and reddening} \label{reddening}

We used the CHAOS project's MODS standard reduction pipeline for calibration and
extraction of the spectrum and for the measurement of the emission line fluxes
\citep[see][]{berg2015, rogers2021}. The modsCCDRed Python programs bias subtract and flatfield the input science, standard star, and calibration lamp images. Sky subtraction and region extraction is performed using the modsIDL reduction pipeline\footnote{\url{http://www.astronomy.ohio-state.edu/MODS/Software/modsIDL/}}, which operates in the XIDL reduction package\footnote{\url{https://www.ucolick.org/~xavier/IDL/}}. Additionally, this pipeline uses the arc lamp images to calculate a wavelength calibration for the 1D spectra and uses the standard star observations to determine the sensitivity function for flux calibration and correction for atmospheric extinction.

The emission line fluxes were measured in the extracted one-dimensional spectrum by fitting Gaussian profiles.  In cases where emission lines were blended, multiple Gaussian profiles were fit to derive the integrated line fluxes.  The errors of the flux measurements were approximated using
\begin{equation}
        \sigma_{\lambda} \approx \sqrt{ {(2\times \sqrt{N}\times \zeta)}^2 + {(0.02\times F_{\lambda})}^2 } ,     
        \label{eq:uncertainty}
\end{equation}
where $N$ is the number of pixels spanning the Gaussian profile fit to the narrow emission lines, and $\zeta$ is the rms noise in the continuum, determined as the average of the rms on each side of an emission line.  For weak lines, the uncertainty is dominated by error from the continuum subtraction, so the rms noise term determines the uncertainty.  For the lines with flux measurements much stronger than the rms noise of the continuum, the error is dominated by the flux calibration.  A minimum flux uncertainty of 2\% was assumed, based on the uncertainties in the standard star measurements \citep{oke90}, and, for the strongest emission lines, it dominates the uncertainty estimate.

The relative intensities of the Balmer lines are used to solve for the reddening using the reddening
law of \citet{ccm89}, assuming $A_{V}=3.1\ E(B-V)$ and
the theoretical case B values from \citet{sh95} interpolated to the temperature
derived from the [\ion{O}{iii}] lines.
We used a minimized $\chi^2$ approach to solve simultaneously for the reddening and underlying absorption based on
the H$\alpha$/H$\beta$, H$\gamma$/H$\beta$, and H$\delta$/H$\beta$ ratios \citep[see][]{os01}.
The solution was consistent with an underlying Balmer absorption with equivalent width of 1.5$^{+0.4}_{-0.4}$ \AA .

The reddening corrected emission line intensities, normalized to H$\beta$ are presented in Table \ref{tab:LineRatios}.  Note that, due to the larger aperture size, higher spectral resolution, and extended wavelength coverage in the red, the LBT spectrum has allowed the detection of many additional faint and red emission lines.  We have reported line intensities for some faint lines that are marginally detected (s/n $<$ 3) as they may be valuable for planning future spectroscopic observations.

The reddening correction can be tested by comparing the
corrected higher numbered Balmer lines to their theoretical values.
The H9 $\lambda$3935 and H10 $\lambda$3798 corrected fluxes are consistent with their theoretical
ratios to H$\beta$ of 0.074 and 0.054, respectively.
The derived values are listed in Table~\ref{tab:LineRatios}.  Note that in Section~\ref{AGC_helium} we
re-derive the reddening (and other parameters) from the combination of H and He lines and that the two derived values are in excellent agreement.  The value derived here, C(H$\beta$) $=$ 0.13 $\pm$ 0.03, while small, is significantly larger than previous determinations (0.04 $\pm$ 0.05 from
the MMT observations and 0.001 from the Keck observations).  At a high Galactic 
latitude ($b = 49.3^{\circ}$), with a corresponding expected foreground
reddening value of C(H$\beta$) $=$ 0.02 \citep{sf2011}, and an extremely low metallicity, this might appear to be unexpectedly large.  However, 
some degree of reddening is expected in any actively star forming region.
The amplitude of the reddening correction does not have a large impact on the derived chemical abundances, but this discrepancy from previous observations is
worth noting.

\begin{table}
\caption{Emission-Line Intensity Ratios Relative to H$\beta$ for AGC~198691}
\label{tab:LineRatios}
\begin{tabular}{lccc}
\hline
{Ion} & {$\lambda$ [\AA]} & {MMT 6.5-m (composite)} & {LBT} \\
\hline
H 14              &  3721     &      \ldots           &  0.023$\pm$0.002 \\
{[O II]}          &  3727.53  &    0.468 $\pm$ 0.025  &  0.429$\pm$0.027  \\
H 13              &  3734     &       \ldots          &  0.029$\pm$0.003 \\
H 12              &  3750.15  &       \ldots          &  0.035$\pm$0.011 \\
H 11              &  3770.63  &       \ldots          &  0.039$\pm$0.011  \\
H 10              &  3797.90  &       \ldots          &  0.046$\pm$0.025  \\
He I              &  3819.64  &        \ldots         &   0.015$\pm$0.004 \\
H 9               &  3835.39  &    0.053 $\pm$ 0.004  &   0.072$\pm$0.023 \\
{[Ne III]}        &  3868.75  &    0.097 $\pm$ 0.006  &   0.114$\pm$0.006 \\
He I + H8         &  3888.65  &    0.168 $\pm$ 0.009  &   0.209$\pm$0.016 \\
{[Ne III]} + H7   &  3969.56  &    0.176 $\pm$ 0.009  &   0.206$\pm$0.022 \\
He I              &  4026.29  &        \ldots         &   0.025$\pm$0.005 \\
{[S II]}          &  4068.60  &        \ldots         &    0.010$\pm$0.004 \\
{[S II]}          &  4076.35  &        \ldots         &   0.008$\pm$0.005 \\
\HD\              &  4101.74  &    0.254 $\pm$ 0.011  &   0.262$\pm$0.012 \\
He I              &  4120.86  &         \ldots        &   0.007$\pm$0.003 \\
He I              &  4143.76  &         \ldots        &  0.005$\pm$0.003 \\
\HG\              &  4340.47  &    0.482 $\pm$ 0.017  &   0.486$\pm$0.016 \\
{[O III]}         &  4363.21  &    0.039 $\pm$ 0.003  &    0.046$\pm$0.004 \\
He I              &  4387.93  &          \ldots       &  0.002$\pm$0.003  \\
He I              &  4471.50  &    0.030 $\pm$ 0.003  &   0.036$\pm$0.003 \\
\HB\              &  4861.33  &    1.000 $\pm$ 0.029  &  1.000$\pm$0.030   \\
He I              &  4921.93  &    0.007 $\pm$ 0.002  &   0.010$\pm$0.002 \\
{[O III]}         &  4958.91  &    0.418 $\pm$ 0.013  &   0.473$\pm$0.022 \\
{[O III]}         &  5006.84  &    1.283 $\pm$ 0.037  &   1.386$\pm$0.065  \\
He I              &  5016     &         \ldots        &  0.024$\pm$0.036  \\
{[N II]}          &  5754.52  &         \ldots        &   0.003$\pm$0.002 \\
He I              &  5875.62  &    0.075 $\pm$ 0.004  &    0.105$\pm$0.007 \\
{[S III]}         &  6312.10  &         \ldots        &  0.008$\pm$0.003 \\
{[N II]}          &  6548.03  &         \ldots        &  0.003$\pm$0.002  \\
\HA\              &  6562.82  &    2.758 $\pm$ 0.137  &  2.766$\pm$0.082    \\
{[N II]}          &  6583.41  &    0.016 $\pm$ 0.002  &   0.017$\pm$0.002 \\
He I              &  6678.15  &    0.023 $\pm$ 0.002  &   0.030$\pm$0.003  \\
{[S II]}          &  6716.47  &    0.037 $\pm$ 0.003  &   0.031$\pm$0.003  \\
{[S II]}          &  6730.85  &    0.031 $\pm$ 0.003  &   0.023$\pm$0.003 \\
He I              &  7065.28  &          \ldots       &   0.025$\pm$0.002  \\
{[Ar III]}        &  7135.78  &          \ldots       &   0.015$\pm$0.002 \\
{[O II]}          &  7319.65  &          \ldots       &   0.008$\pm$0.002  \\
{[O II]}          &  7330.16  &          \ldots       &   0.006$\pm$0.002 \\
{[Ar III]}        &  7750.60  &          \ldots       &   0.011$\pm$0.005  \\
P 11              &  8862.79  &          \ldots       &   0.019$\pm$0.010  \\
P 10              &  9014.91  &          \ldots       &  0.023$\pm$0.003 \\
{[S III]}         &  9068.90  &           \ldots      &   0.028$\pm$0.004 \\
P 9               &  9229.02  &          \ldots       &  0.020$\pm$0.003 \\
{[S III]}         &  9531.00  &           \ldots      &  0.066$\pm$0.010 \\
P 8               &  9546.20  &          \ldots       &    0.040$\pm$0.011 \\
\\
\hline
\\
c(\HB)           &           &    0.040 $\pm$ 0.053    & 0.137 $\pm$ 0.029   \\
EW(\HB)          &           &    64.3 \AA             &  78.2 \AA \\
F(\HB)\tablefootnote{line flux of H$\beta$ in units of 10$^{-16}$ erg s$^{-1}$ cm$^{-2}$}                 &           &    9.25 $\pm$ 0.19      & 9.65 $\pm$ 0.19 \\
\hline
\end{tabular}
\end{table}

\subsection{The He~I $\lambda$5876 emission line} \label{sec:5876}

Due to AGC~198691's velocity (514 km s$^{-1}$) and the 1$^{\prime\prime}$ width of the LBT/MODS slit, the He~I $\lambda$5876 emission line partially overlapped with the Na~I $\lambda$5889.95 night sky emission line.
The potential interference of the Na~I D lines with He~I $\lambda$5876 and the
role that it plays in deriving He abundances was highlighted by \citet{dk85} and \citet{ds86}.  The strong Na~I atmospheric emission can complicate the subtraction of the night sky from the target spectrum, but there is also the possibility of absorption from the intervening Galactic neutral interstellar medium.  For these reasons, \citet{os04} proposed screening out targets used for the determination of $Y_p$ in the heliocentric velocity ranges of 728 $\pm$ 100 and 1032 $\pm$ 100 km s$^{-1}$.  Of course, those ranges can be affected by the spectral resolution of the observations.  Note that this screen is not applied uniformly across all determinations of $Y_p$ in the literature.  AGC~198691, with a radial velocity of 514 km s$^{-1}$, would appear to be safely outside of this range, but our LBT spectrum reveals otherwise.  Thus, a more conservative range of $\pm$ 200 km s$^{-1}$ may be appropriate.

Sky subtraction in the modsIDL pipeline is accomplished by fitting night sky emission lines with a B-spline in the dispersion direction and a low-order polynomial in the slit direction \citep{kelson03}.  
He~I $\lambda$5876 is the strongest optical helium emission line, and it carries corresponding importance in our analysis.  As a result, without He~I $\lambda$5876, the precision of our helium abundance determination for AGC~198691 would be significantly reduced, and the reliability of the results suspect.  Therefore, extracting a reliable measurement for He~I $\lambda$5876 was a high priority.  Luckily, only the higher-wavelength side of the emission line was clipped by the night sky line, which assisted in assessing the quality of the sky subtraction.  

To extract a reliable He~I $\lambda$5876, the extraction window and sky subtraction regions were chosen to optimize the extraction of He~I $\lambda$5876, as well as the rest the spectrum.  The extraction was chosen to optimize both the signal-to-noise measure for the line, as well as the symmetry of the line profile.  As mentioned in \S \ref{LBT}, a 22 pixel extraction window was employed, along with symmetrically chosen sky regions.  
The measured He~I $\lambda$5876/H$\beta$ flux ratio is reported in the next section (\S \ref{HeFluxes}) in Table \ref{table:Fluxes}, along with the other He and H lines used in our helium abundance analysis.  

\subsection{H and He Emission Line Fluxes and Equivalent Widths} \label{HeFluxes}

In Table \ref{table:Fluxes}, we present the emission line fluxes, equivalent widths, and their associated uncertainties for all of the H and He recombination lines necessary for the helium abundance analysis of AGC~198691. 
The uncertainties on the flux measurements were approximated using equation \ref{eq:uncertainty}.

\begin{table}
\centering
\caption{H and He Emission Line Fluxes and Equivalent Widths for AGC~198691}
\begin{tabular}{lcc}
\hline\hline
Ion 		        	& F($\lambda$)/F(H$\beta$)  &  W($\lambda$) [\AA]	\\	
\hline														
H11 $\lambda$3771	&	0.0145	$\pm$	0.0051	&	0.74	\\
H10 $\lambda$3798	&	0.0223	$\pm$	0.0036	&	1.15	\\
H9 $\lambda$3835	&	0.0393	$\pm$	0.0053	&	1.96	\\
He~I+H8 $\lambda$3889	&	0.1515	$\pm$	0.0054	&	7.71	\\
He~I $\lambda$4026	&	0.0125	$\pm$	0.0036	&	0.65	\\
H$\delta$ $\lambda$4101	&	0.2200	$\pm$	0.0075	&	12.27	\\
H$\gamma$ $\lambda$4340	&	0.4457	$\pm$	0.0134	&	28.26	\\
He I $\lambda$4471	&	0.0307	$\pm$	0.0037	&	2.05	\\
H$\beta$ $\lambda$4861	&	1.0000	$\pm$	0.0290	&	77.37	\\
He~I $\lambda$5015	&	0.0215	$\pm$	0.0033	&	1.76	\\
He~I $\lambda$5876	&	0.1114	$\pm$	0.0045	&	12.82	\\
H$\alpha$ $\lambda$6563	&	3.1031	$\pm$	0.0890	&	435.0	\\
He~I $\lambda$6678	&	0.0322	$\pm$	0.0019	&	4.77	\\
He~I $\lambda$7065	&	0.0249	$\pm$	0.0035	&	3.98	\\
P10 $\lambda$9015	&	0.0209	$\pm$	0.0034	&	5.79	\\
P9 $\lambda$9229	&	0.0227	$\pm$	0.0040	&	7.27	\\
P8 $\lambda$9546	&	0.0481	$\pm$	0.0141	&	14.69	\\
\hline	
 		        	& F($\lambda$)/F(P$\gamma$)  &  W($\lambda$) [\AA]	\\	
\hline	
He I $\lambda$10830	&	4.61053	$\pm$	0.4232	&	58.20	\\
P$\gamma$ $\lambda$10940	&	1.00	$\pm$	0.73	&	5.29	\\
\hline
\hline
\multicolumn{3}{l}{F(H$\beta$) = (0.9539 $\pm$ 0.0196) $\times$ 10$^{-15}$ erg s$^{-1}$ cm$^{-2}$} \\
\hline
\end{tabular}
\label{table:Fluxes}
\end{table}

The \10830 emission line for AGC~198691 was observed by \citet{hcpb} using Keck NIRES.  It is reported as J0943+3326 after being independently identified by \citet{hsyu2018}.  Given the utility of this emission line for helium abundance determinations \citep{itg14,AOS4}, we are very thankful to have this observation from \citet{hcpb} for our analysis.

\section{Results} \label{Results}

\subsection{Physical Conditions and Elemental Abundances}  \label{abundances}

\begin{table}
\caption{Electron Temperatures \& Densities and Ionic \& Elemental Abundances}
\begin{tabular}{lcc}
\hline\hline
{Property}                                          &{MMT 6.5-m (composite)} & {LBT} \\
\hline
\noindent $T_{e}$ (\ion{O}{III}) [K]         & 19130 $\pm$ 800 & 
19680 $\pm$ 790 \\
\noindent $t_{2}$ (Inferred) [K]         & 15120 $\pm$ 560 &  15290 $\pm$ 550 \\
\noindent $t(S^{2+})$ (Inferred) [K]         & \ldots & 
16340 $\pm$ 830 \\
\noindent $n_{e}$ (\ion{S}{II}) [cm$^{-3}$]  & 270 $\pm$ 200   &  
60 $\pm$ 220 \\
\\
\noindent O$^{+}$/H$^{+}$ [$\times$10$^{6}$]        & 3.11 $\pm$ 0.24 &  
3.35 $\pm$ 0.43 \\
\noindent O$^{++}$/H$^{+}$ [$\times$10$^{6}$]       & 7.35 $\pm$ 0.51 &  
8.00 $\pm$ 0.71 \\
\noindent O/H [$\times$10$^{6}$]                    & 10.46 $\pm$ 0.68 &  
11.35 $\pm$ 0.82  \\
\noindent 12+log(O/H)                               & 7.02 $\pm$ 0.03 &  
7.06 $\pm$ 0.03  \\
\\
\noindent N$^{+}$/H$^{+}$ [$\times$10$^{7}$]        & 0.95 $\pm$ 0.12 &  
1.39 $\pm$ 0.20  \\
\noindent ICF (N)                                   & 3.37 $\pm$ 0.16 &  
3.39 $\pm$ 0.50  \\
\noindent 12+log(N/H)                               & 5.51 $\pm$ 0.06 &  
5.67 $\pm$ 0.08  \\
\noindent log(N/O)                                  & -1.51 $\pm$ 0.06 &  
-1.38 $\pm$ 0.06  \\
\\
\noindent Ne$^{++}$/H$^{+}$ [$\times$10$^{6}$]      & 1.22 $\pm$ 0.11 &  
1.48 $\pm$ 0.16  \\
\noindent ICF (Ne)                                  & 1.42 $\pm$ 0.03 &  
1.42 $\pm$ 0.16  \\
\noindent 12+log(Ne/H)                              & 6.24 $\pm$ 0.04 &  
6.32 $\pm$ 0.06  \\
\noindent log(Ne/O)                                 & -0.78 $\pm$ 0.02 &  
-0.73 $\pm$ 0.03  \\
\\
\noindent S$^{+}$/H$^{+}$ [$\times$10$^{7}$]      & \ldots &  
0.55 $\pm$ 0.05  \\
\noindent S$^{++}$/H$^{+}$ [$\times$10$^{7}$]      & \ldots &  
1.84 $\pm$ 0.32  \\
\noindent ICF (S)                                  & \ldots &  
1.40 $\pm$ 0.14  \\
\noindent 12+log(S/H)                              & \ldots &  
5.52 $\pm$ 0.07  \\
\noindent log(S/O)                                 & \ldots &  
-1.53 $\pm$ 0.07  \\
\\
\noindent Ar$^{++}$/H$^{+}$ [$\times$10$^{8}$]      & \ldots &  
4.59 $\pm$ 0.76  \\
\noindent ICF (Ar)                                  & \ldots &  
1.09 $\pm$ 0.11  \\
\noindent 12+log(Ar/H)                              & \ldots &  
4.70 $\pm$ 0.08  \\
\noindent log(Ar/O)                                 & \ldots &  
-2.35 $\pm$ 0.08  \\
\hline
\end{tabular}
\label{tab:Abundances}
\end{table}

\subsubsection{Electron Temperature and Density Determinations}\label{sec:phys}

For the purpose of deriving nebular abundances,
we adopt a three zone approximation, where $t_2$, $t(S^{2+})$, and $t_3$ are the electron temperatures (in units of $10^{4}$ K) in the low, intermediate, and high ionization zones respectively.
For the high ionization zone, the [\ion{O}{iii}] I($\lambda\lambda$4959,5007)/I($\lambda$4363)
ratio was used to derive a temperature.
We use \textsc{PyNeb} \citep{luridiana12,luridiana15} to calculate the best-fit electron temperature from the measured auroral-to-nebular line flux ratio and using a density of $n_{\rm e}=10^2$ cm$^{-3}$.  
The derived temperature is given in Table~\ref{tab:Abundances}.
We derive a relatively high temperature of 19,680 $\pm$ 790 K from the LBT/MODS spectrum, as expected for
a low metallicity \ion{H}{ii} region and in good agreement with the temperature obtained from the MMT composite spectrum of \citet{hirschauer2016}.

Because neither the [\ion{O}{ii}] $\lambda\lambda$7320,7320 lines nor the
[\ion{N}{ii}] $\lambda$5755 line were detected at high significance, we cannot
derive a temperature for the low ionization zone directly and, therefore,
need to assume a temperature. We used the T$_e$-T$_e$ relations proposed by \citet{pagel92}, based on the photoionization modeling of \citet{stasinska90} to determine the low ionization zone temperature:
\begin{equation}
        {t_{2}}^{-1} = 0.5(({t_{3}})^{-1} + 0.8).
        \end{equation}
The [\ion{S}{iii}] $\lambda$6312 emission line was only detected at the 2-$\sigma$ level,
so for the temperature in the [\ion{S}{iii}] zone we assumed a temperature based on
the relationship derived by \citet{garnett92}:
\begin{equation}
        t(S^{+2}) = 0.83(t_{3}) + 0.17.
        \end{equation}
To account for the uncertainties/shortcomings of T$_e$-T$_e$ relations in accounting for real electron temperature variations, the uncertainty in the inferred $t_{2}$ and $t(S^{+2})$ is calculated as the addition of the uncertainty on $t_{3}$ and 500 K in quadrature. 
The different ionization zone temperatures are tabulated in Table~\ref{tab:Abundances}.

[\ion{S}{ii}] $\lambda\lambda$6717,6731 and $t_2$
were used to determine the electron density. This density is consistent with the low density limit, or $n_{\rm e}$ $<$ 10$^2$ cm$^{-3}$. As such, we assume $n_{\rm e}=10^2$ cm$^{-3}$ for all abundance calculations.


\subsubsection{Oxygen and Nitrogen Abundance Determinations}\label{sec:oxygen}

We determine oxygen abundances based on our estimated ionization zone electron temperatures.
Ionic abundances were calculated with:
\begin{equation}
        {\frac{N(X^{i})}{N(H^{+})}\ } = {\frac{I_{\lambda(i)}}{I_{H\beta}}\ } {\frac{j_{H\beta}}{j_{\lambda(i)}}\ },
        \label{eq:Nfrac}
\end{equation}
where the $I$ values are the relative emission line intensities and the $j$ values are the
volume emissivities for the individual emission lines. The emissivities are functions of electron temperature within the ionization zone containing the specific ionic species. For the low-ionization species of N$^+$, O$^+$, and S$^+$, we use $t_{2}$; for high-ionization species such as O$^{2+}$ and Ne$^{2+}$, we use the directly measured temperature from [\ion{O}{iii}] $\lambda$4363, or $t_{3}$; for S$^{2+}$ and Ar$^{2+}$, we apply the inferred  $t(S^{2+})$. To determine the abundances, we use \textsc{PyNeb}, assuming a five-level atom model \citep{derob1987}, the updated atomic data used in \citet{berg2015}, the electron temperature of a given ionization zone, and assuming an electron density of 10$^{2}$ cm$^{-3}$. 
The total oxygen abundance, O/H, is the sum of O$^+$/H$^+$ and O$^{++}$/H$^+$.

The oxygen abundance determination is given in Table~\ref{tab:Abundances}.
We derive an oxygen abundance of 12 + log(O/H) = 7.06 $\pm$ 0.03.  
This value agrees well with the value derived by
\citet{hirschauer2016} of 12 + log(O/H) = 7.02 $\pm$ 0.03.

As a result, the oxygen abundance in AGC~198691 is still one of the lowest oxygen abundances ever derived for an \ion{H}{ii} region. 
The oxygen abundance in AGC~198691 is less than that of 
I~Zw~18 \citep[with 12 + log(O/H) = 7.17 $\pm$ 0.04;][]{skillman93, izotov99}, SBS~0335-052E \citep[log(O/H) = 7.33 $\pm$ 0.01;][]{izotov97}, SBS~0335-052W \citep[log(O/H) = 7.12 $\pm$ 0.03;][]{izotov05},
and DDO~68 \citep[log(O/H) = 7.14 $\pm$ 0.03;][]{pustilnik05, izotov07b}. However, it is formally less metal-poor than the current record holder
J0811+4730 with an oxygen
abundance of 12 + log(O/H) $=$ 6.98 $\pm$ 0.02 \citep{itgl18}.

We derive the N/O abundance ratio from the [\ion{O}{ii}]$\lambda$3727/[\ion{N}{ii}]$\lambda$6584 ratio and
assume N/O = N$^{+}$/O$^{+}$ \citep{peimbert69}.
\cite{nava06} have investigated the validity of this assumption.
They concluded that, although it could be improved upon with modern photoionization models,
it is valid to within about 10\%.
The nitrogen to oxygen relative abundances are given in Table~\ref{tab:Abundances}; the value for log(N/O) determined from the LBT/MODS spectrum is $-$1.38 $\pm$ 0.06.

The newly derived value of N/O in AGC~198691 is higher than 
the previous AGC~198691 determination from \citet{hirschauer2016}.
It is also significantly
higher than the very narrow plateau at $-$1.60 $\pm$ 0.02 which \citet{izotov99} highlighted
in their study of very low metallicity emission line galaxies. 
However, it is comparable to the value derived for Leo~P ($-$1.36 $\pm$ 0.04)
and the low metallicity isolated dwarf irregular galaxies observed by \citet{vanzee2006}.

\subsubsection{Neon, Sulfur, and Argon Abundances}

To estimate the neon abundance, we assume that Ne/O =  Ne$^{++}$/O$^{++}$
\citep{peimbert69}.
The neon to oxygen relative abundance determinations are given in Table~\ref{tab:Abundances}.
We derive log(Ne/O) = $-$0.73 $\pm$ 0.03. 
This is in good agreement with the value observed in Leo P of $-$0.76 $\pm$ 0.03
\citep{skillman2013} and typical of values derived in other XMP galaxies \citep{izotov99, itg12}.

To determine the sulfur and argon abundances, for direct comparison, we adopt ionization
correction factors (ICF) from the literature. For sulfur, we use the ICF from \citet{thuan95} generated from \ion{H}{ii} region photoionization models to correct for the unobserved S$^{+3}$. To account for the Ar$^{+2}$, and Ar$^{+4}$ states, we use [\ion{Ar}{iii}]$\lambda$7135 in conjunction with the low-metallicity ICF of \citet{iz06}. In both cases, we assume a 10\% uncertainty when applying these ICFs. 
For log(S/O) we obtain $-$1.53 $\pm$ 0.07.  This relative abundance agrees with the value found in Leo~P of $-$1.49 $\pm$ 0.07, and is comparable to the average of values reported for other XMP galaxies by \citet{izotov99, itg12}.
For log(Ar/O) we obtain $-$2.35 $\pm$ 0.08.
This is significantly lower than the value found for Leo~P of $-$2.00 
$\pm$ 0.09 but consistent with the range of log(Ar/O) measured in other XMP galaxies 
found by \citet{izotov99, itg12}.

Overall the abundances of neon, sulfur, and argon are consistent with
the trends for $\alpha$ elements seen in low metallicity dwarfs, i.e.,
they appear to be constant as a function of metallicity \citep[e.g.,][]{thuan95}. This is easiest to understand as a result of a nearly universal stellar 
initial mass function and no strong dependence of nucleosynthetic yields on metallicity.

\subsection{The Helium Abundance}  \label{He}

\subsubsection{Methodology} \label{Model}

H~II regions are emission nebulae photoionized by one or more massive stars (O \& B stars).  Correspondingly, their spectra include nebular emission lines overlaid on a stellar continuum.  In seeking the helium abundance, $y^{+}=\frac{n(He~II)}{n(H~II)}$, we model the primary emission region using a radiative transfer model based on the case B approximation (optically thick to Lyman lines).  For hydrogen and helium, recombination emission dominates, but collisional excitation also contributes, with hydrogen collisional excitation from the small fraction of neutral hydrogen in the ionized region, $\xi$.  Self-absorption and re-emission of photons within the nebula also occurs.  The radiative transfer model assumes a uniform H~II region with average electron density, $n_e$, and average electron temperature, $T_e$, as well as optical depth, $\tau$.  The stellar continuum juxtaposes absorption features under the nebular helium and hydrogen emission lines, including separate parameters for the Balmer and Paschen lines \citep{AOS5}, $a_{He}$, $a_H$, and $a_P$, respectively.  Dust along the line of sight also scatters the emitted photons (interstellar reddening, $C(H\beta)$).  

We determine the \he4 abundance in an individual H~II region based on a Markov Chain Monte Carlo (MCMC) analysis. The MCMC method is an algorithmic procedure for sampling from a statistical distribution \citep{mar,met}.  
We define a $\chi^2$ distribution from the difference between
flux ratios, 
\beq
\chi^2 = \sum_{\lambda} \frac{(\frac{F(\lambda)}{F(H\beta|P\gamma)} - {\frac{F(\lambda)}{F(H\beta|P\gamma)}}_{meas})^2} { \sigma(\lambda)^2},
\label{eq:X2}
\eeq
where the emission line fluxes, $F(\lambda)$, are measured and calculated for a set of H and He lines, and $\sigma(\lambda)$ is the measured uncertainty in the flux ratio at each wavelength. 
The optical/near-IR emission line fluxes from the LBT/MODS spectrum are calculated relative to H$\beta$, while the infrared flux \10830 from \citet{hcpb} is calculated relative to the IR Paschen line, P$\gamma$.  Thus, by $F(H\beta|P\gamma)$, we mean $F(H\beta)$ for all lines other than \10830, and $F(P\gamma)$ for the latter. 

For this work seven helium line ratios are employed:  $\lambda\lambda$4026, 4471, 5015, 5876, 6678, and 7065, relative to H$\beta$, and $\lambda$10830, relative to P$\gamma$.  Nine hydrogen line ratios are employed:  H$\alpha$, H$\gamma$, H$\delta$, H9, H10, H11, P8, P9, and P10, relative to H$\beta$.  Finally, the blended line \3889 + H8, relative to H$\beta$, is also employed. Note that four lines ($\lambda$4922, H12, P11, and P12) considered in \citet{AOS5} are not used here as they are too faint to contribute, as can be seen from Table~\ref{tab:LineRatios}.

These observed line ratios are used to fit nine model parameters, as introduced above \citep{AOS5}: $T_e$, $n_e$, $\tau$, $C(H\beta)$, $a_{He}$, $a_H$, $a_P$, $\xi$, and $y^+$.  Thus, we are left with 8 degrees of freedom, corresponding to a total of 17 observed line ratios and 9 parameters to fit. 

The MCMC scans of our 9-dimensional parameter space map out the $\chi^2$ distribution given above.  We conduct a frequentist analysis, and the $\chi^2$ is minimized to determine the best-fit solution for the nine physical parameters, including $y^+$, as well as determining the ``goodness-of-fit''.  Uncertainties in each quantity are estimated by calculating a 1D marginalized likelihood and finding the 68\% confidence interval from the increase in the $\chi^2$ from the minimum.  
The model equations for the helium and hydrogen flux ratios are given in full and our entire model and the data employed are detailed in the extensive appendices in \citet{AOS5}.

\subsubsection{The Helium Abundance}  \label{AGC_helium}

Using our expanded and updated model, as outlined in \S \ref{Model}, and the new observations and measurements discussed in \S \ref{AGC}, AGC~198691 was analyzed using our MCMC analysis \citep{AOS2, AOS5}.  The best-fit model parameter values and uncertainties are given in Table \ref{table:AGC}.  

\begin{table}
\centering
\caption{Physical conditions of AGC~198691 from H and He only}
\begin{tabular}{lc}
\hline\hline
Emission lines          &  17            \\
Free Parameters         &  9             \\
degrees of freedom      &  8             \\
95\% CL $\chi^2$        &  15.51         \\
\hline
He$^+$/H$^+$		    &  0.0776$^{+0.0057}_{-0.0032}$     \\
n$_e$ [cm$^{-3}$]	    &  231$^{+207}_{-159}$              \\
a$_{He}$ [\AA]		    &  0.30$^{+0.17}_{-0.16}$           \\
$\tau$				    &  0.00$^{+0.55}_{-0.00}$           \\
T$_e$ [K]			    &  14,900$^{+2000}_{-2600}$         \\
C(H$\beta$)			    &  0.13$^{+0.02}_{-0.07}$           \\
a$_H$ [\AA]			    &  1.69$^{+0.21}_{-0.19}$           \\
a$_P$ [\AA]             &  0.97$^{+0.60}_{-0.97}$           \\
$\xi$ $\times$ 10$^4$   &  0$^{+402}_{-0}$                \\
\hline
$\chi^2$			    &  6.28                             \\
p-value                 &  62\%                             \\
\hline
O/H $\times$ 10$^5$     &  1.1 $\pm$ 0.1 \\
Y                       &  0.2367  $\pm$   0.0132           \\
\hline
\end{tabular}
\label{table:AGC}
\end{table}

The helium abundance uncertainty is significantly larger than was observed for Leo~P ($y^{+}$ = 0.0776$^{+0.0057}_{-0.0032}$ vs. 0.0823$^{+0.0025}_{-0.0018}$) \citep{AOS5}.  Compared to Leo~P, the larger relative uncertainties for the added hydrogen lines (and in particular compared to the three strongest hydrogen line ratios), limits their ability to constrain the solution.  These larger relative uncertainties are due to the lower signal-to-noise for AGC~198691's spectrum, compared to Leo~P's.  AGC~198691 is a faint object (F(H$\beta$) is 3.5 times weaker), and this translates into lower signal-to-noise and increased uncertainties on the flux measurements.  Similarly, the \10830 observation from \citet{hcpb} carries a larger relative uncertainty than the corresponding observation for Leo~P (9.18\% vs. 3.18\%).  
The helium mass fraction (used for the regression in \S \ref{Yp}) is Y = 0.2367 $\pm$ 0.0132.  

The reddening determined by our helium abundance model, 0.13$^{+0.02}_{-0.07}$, is in excellent agreement with the 0.14 $\pm$ 0.03 value found in our abundance analysis in Section \ref {reddening}.  These are, of course, not fully independent determinations, but, since the helium abundance model involves many more model parameters and is based on only the helium and hydrogen emission lines, the agreement of these results was not assured, and their consistency adds support to the robustness of the result.  Similar to the value reported in \ref{reddening}, the value found here is also in agreement with the previous determination from \citet{hirschauer2016} (0.04 $\pm$ 0.05), albeit primarily due to the relative large uncertainties on both values.  

The physical solution looks reliable, with most best-fit physical parameters not constrained by their zero lower bound, and most of the parameter ranges (68\% CI) also separated from zero.  Along with our recent Leo~P analysis \citep{AOS5}, this helps bolster a conclusion from our first work incorporating \10830 \citep{AOS4}.  In our more recent analyses, employing more emission lines to better determine and constrain the physical model parameters, the best-fit parameter values reach their lower limits less frequently than in our preceding analyses \citep{AOS3, AOPS}.  Correspondingly, the solutions more reliably report unambiguously valid physical values.  In addition to the solution's overall qualitative reliability, it should be noted that parameters reaching their physical lower limits reduces the uncertainties determined for all parameters (see \citet{AOS2} and 1D marginalization for constructing parameter confidence intervals), which could then be viewed as underestimated.  It is for this reason that it is important to assess the reliability of the physical solution based on its full parameter solution, as well as its best-fit $\chi^2$ value.  

Our physical solution for AGC~198691 fits 9 parameters using 17 emission line ratios. As indicated in Table \ref{table:AGC}, the total $\chi^2$ for the best-fit solution was 6.22, for eight degree of freedoms, corresponding to a p-value of 62\%.  
Note that the solutions for $T_e$ \& $n_e$ found here, using our solution for $y^+$, are consistent with those listed in Table~\ref{tab:Abundances}, derived from \ion{O}{iii} and \ion{S}{ii}, respectively.  It should also be noted that $T_e(\ion{He}{II})$ is lower than $T_e(\ion{O}{III})$, as expected.   
Table \ref{table:chi} provides a full comparison of the measured fluxes and the model predictions for AGC~198691, broken down by line.  The relative uncertainty for each measured flux is also included.  The stronger lines dominate the $\chi^2$ minimization and best-fit solution, due to their lower relative uncertainties. However, the weaker lines, appropriately weighted by their larger uncertainties, still contribute to constrain the best-fit solution and parameter uncertainties.  Of particular note, He~I $\lambda$5876 agrees well with its model-predicted flux (i.e., it agrees well with the other lines in their parameter determination).  Its small $\chi^2$ contribution (the second smallest of the optical helium lines) provides support for the robustness of its extraction from the night-sky emission line and its resulting reliability.  

\begin{table}
\centering
\caption{Comparison of the measured flux ratio to the model prediction by line for the best-fit solution for AGC~198691.}
\begin{tabular}{lccccc}
\hline\hline
Emission Line 		        	& $(\frac{\textrm{F}(\lambda)}{\textrm{F(H}\beta)})_{\textrm{meas}}$  &  $\sigma$    & \%  & $(\frac{\textrm{F}(\lambda)}{\textrm{F(H}\beta)})_{\textrm{mod}}$   &   $\chi^2$    \\	
\hline											
\multicolumn{6}{l}{\textit{Helium}:}    \\											
He~I $\lambda$4026	&	0.0125	&	0.0036	&	28.57	&	0.0104	&	0.333	\\
He I $\lambda$4471	&	0.0307	&	0.0037	&	12.07	&	0.0324	&	0.211	\\
He~I $\lambda$5015	&	0.0215	&	0.0033	&	15.51	&	0.0231	&	0.210	\\
He~I $\lambda$5876	&	0.1114	&	0.0045	&	4.04	&	0.1095	&	0.187	\\
He~I $\lambda$6678	&	0.0322	&	0.0019	&	6.03	&	0.0317	&	0.078	\\
He~I $\lambda$7065	&	0.0249	&	0.0035	&	14.00	&	0.0274	&	0.500	\\
\multicolumn{6}{l}{\textit{Hydrogen}:}	\\										
H11 $\lambda$3771	&	0.0145	&	0.0051	&	35.10	&	0.0117	&	0.320	\\
H10 $\lambda$3798	&	0.0223	&	0.0036	&	15.93	&	0.0229	&	0.025	\\
H9 $\lambda$3835	&	0.0393	&	0.0053	&	13.54	&	0.0396	&	0.003	\\
H$\delta$ $\lambda$4101	&	0.2200	&	0.0075	&	3.42	&	0.2218	&	0.056	\\
H$\gamma$ $\lambda$4340	&	0.4457	&	0.0134	&	3.01	&	0.4349	&	0.658	\\
H$\alpha$ $\lambda$6563	&	3.1031	&	0.0890	&	2.87	&	3.1096	&	0.005	\\
P10 $\lambda$9015	&	0.0209	&	0.0034	&	16.09	&	0.0182	&	0.624	\\
P9 $\lambda$9229	&	0.0227	&	0.0040	&	17.59	&	0.0270	&	1.150	\\
P8 $\lambda$9546	&	0.0481	&	0.0141	&	29.22	&	0.0402	&	0.314	\\
\multicolumn{6}{l}{\textit{Blended He}+\textit{H}:}	\\										
He~I+H8 $\lambda$3889	&	0.1515	&	0.0054	&	3.56	&	0.1532	&	0.101	\\
\hline											
 		        	& $(\frac{\textrm{F}(\lambda)}{\textrm{F(P}\gamma)})_{\textrm{meas}}$  &  $\sigma$    & \%  & $(\frac{\textrm{F}(\lambda)}{\textrm{F(P}\gamma)})_{\textrm{mod}}$   &   $\chi^2$    \\
\hline											
He I $\lambda$10830	&	4.6105	&	0.4232	&	9.18	&	4.5763	&	0.007	\\
\hline											
\end{tabular}
\label{table:chi}
\end{table}

\subsubsection{Relevance for the Primordial Helium Abundance} \label{Yp}

A regression of Y, the helium mass fraction, versus O/H, the oxygen abundance, from nearby galaxies, is used to extrapolate to the primordial value\footnote{This work takes $Z=20(O/H)$ such that $Y=\frac{4y(1-20(O/H))}{1+4y}$}.  Besides AGC~198691, the O/H values are taken directly from \citet{its07}, except for Leo~P, where the value is taken from \citet{skillman2013}.  

\begin{figure}
\includegraphics[width=\columnwidth]{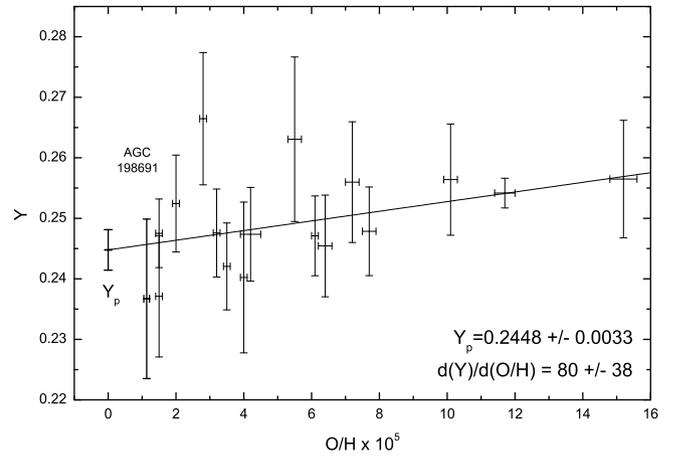}
\caption{
Helium abundance (mass fraction) versus oxygen to hydrogen ratio regression calculating the primordial helium abundance.  The added point, AGC~198691, based on the analysis of this work, is shown as bold.  
}
\label{Y_OH}
\end{figure}

The relevant values for the objects other than AGC 198691 used in the regression are given \citet{AOS5}.  The regression is based on the results from \citet{AOS4}, combined with the results for Leo~P from \citet{AOS5} and those for AGC~198691 from this work.  There are 15 objects in the qualifying dataset of \citet{AOS4}, Leo~P was added in \citet{AOS5}, and AGC~198691 is added to the previous results for those 16.  Regression based on those 17 points yields
\beq
Y_p = 0.2448 \pm 0.0033,
\label{eq:Yp}
\eeq
with a slope of 80 $\pm$ 38 and a total $\chi^2$ of 8.1.  The result is shown in Fig.~\ref{Y_OH}. This result for $Y_{p}$ agrees well with the SBBN value of $Y_p = 0.2469 \pm 0.0002$ \citep{fields2020}, based on the Planck determined baryon density \citep{planck18}.  Eq.\:(\ref{eq:Yp}) also agrees well with the SBBN-independent, direct Planck estimation of $Y_p = 0.239 \pm 0.013$ \citep{planck18}, using temperature, polarization, and lensing data, which is not surprising given the relatively large uncertainty on the CMB measurement. 

\citet{AOS4} determined $Y_p = 0.2449 \pm 0.0040$, with a slope of 79 $\pm$ 43, and the addition of Leo~P increased the precision to $Y_p = 0.2453 \pm 0.0034$, with a slope of 75 $\pm$ 39 \citep{AOS5}.  Since the result given in Eq.\:(\ref{eq:Yp}) is based on the same dataset, with only one additional point (AGC~198691), it is not surprising that the results are so similar.  Furthermore, AGC~198691's larger uncertainty, due to its faintness, limits its ability to constrain the regression.  Our result given in Eq.~(\ref{eq:Yp}) is consistent with other recent determinations \citep{vpps,ftdt2,hcpb,kurichin2021}.

\section{Conclusions} \label{Conclusion}

We have presented a new spectroscopic observation of the XMP dwarf galaxy, AGC~198691, taken by the LBT's MODS instrument.  From this, we have reported an expanded chemical abundance analysis.  Here we summarize our main findings and conclusions:  

\begin{enumerate}
    \item The higher resolution and broader wavelength coverage afforded by LBT/MODS allowed for the first estimation of the sulfur, argon, and helium abundances for AGC~198691.  
    \item The oxygen abundance is in good agreement with the first determination for AGC~198691, as reported in \citet{hirschauer2016}, rather than the slightly higher value found by \citet{hcpb}.  
    \item This confirms AGC~198691 as one of the lowest metallicity galaxies ever discovered.  Such XMP galaxies play a crucial role in studying and constraining models of chemical evolution and star formation and evolution in nearly pristine gas. 
    \item The abundances were broadly consistent with previous determinations \citep{hirschauer2016}, though the spectral measurements showed an increased amount of reddening, with corresponding effects on the abundance results.  
    \item Aided by the higher Balmer \& Paschen lines enabled by LBT/MODS and a \10830 observation from \citet{hcpb}, we were able to apply our recently expanded and improved model \citep{AOS5} to determine the helium abundance for AGC~198691, with a well-determined full model parameter solution.
    \item The faintness of AGC~198691 resulted in relatively large flux uncertainties and a corresponding large uncertainty on its helium abundance.  As a result, even though AGC~198691 agreed well with its predicted helium abundance, based on our most recent regression analysis \citep{AOS5}, its inclusion only very slightly reduced the uncertainty on the primordial helium abundance, Y$_\textrm{p}$.  
    \item This highlights that faint objects, even if extremely low metallicity, have limited ability to improve the determination of the primordial helium abundance.  Therefore, higher surface brightness objects should be prioritized, with both metallicity and surface brightness balanced as important selection criteria.  Specifically, targets with EW(H$\beta$ $>$ 200) should be highest priority.  
    \item As evidenced by the robust results from our expanded helium abundance model, future observations of the most promising XMP galaxies, with high quality, high resolution, and broad wavelength coverage spectra, provide a uniquely promising route for improving the primordial helium abundance determination from observations of galaxies.  
\end{enumerate}

\section*{Acknowledgements}

The work of KAO is supported in part by DOE grant DE-SC0011842.  EDS is grateful for partial support from the University of Minnesota.  
EA benefited greatly from three visits during sabbatical to the University of Minnesota and is grateful to the University of Minnesota and the William I. Fine Theoretical Physics Institute for the support.  

This paper uses data taken with the MODS spectrographs built with funding from NSF grant AST-9987045 and the NSF Telescope System Instrumentation Program (TSIP), with additional funds from the Ohio Board of Regents and the Ohio State University Office of Research.
This paper made use of the modsIDL spectral data reduction pipeline developed in part with funds provided by NSF Grant AST-1108693.
This work was based in part on observations made with the Large Binocular Telescope (LBT). The LBT is an international collaboration among institutions in the United States, Italy and Germany. The LBT Corporation partners are: the University of Arizona on behalf of the Arizona university system; the Istituto Nazionale di Astrofisica, Italy; the LBT Beteiligungsgesellschaft, Germany, representing the Max Planck Society, the Astrophysical Institute Potsdam, and Heidelberg University; the Ohio State University; and the Research Corporation, on behalf of the University of Notre Dame, the University of Minnesota, and the University of Virginia.
This research has made use of the NASA/IPAC Extragalactic Database (NED) which is operated by the
Jet Propulsion Laboratory, California Institute of Technology, under contract with the
National Aeronautics and Space Administration and the NASA Astrophysics Data System (ADS).


\bibliographystyle{mnras}
\bibliography{AGC198691}

\begin{thebibliography}{}
\makeatletter
\relax
\def\mn@urlcharsother{\let\do\@makeother \do\$\do\&\do\#\do\^\do\_\do\%\do\~}
\def\mn@doi{\begingroup\mn@urlcharsother \@ifnextchar [ {\mn@doi@}
  {\mn@doi@[]}}
\def\mn@doi@[#1]#2{\def\@tempa{#1}\ifx\@tempa\@empty \href
  {http://dx.doi.org/#2} {doi:#2}\else \href {http://dx.doi.org/#2} {#1}\fi
  \endgroup}
\def\mn@eprint#1#2{\mn@eprint@#1:#2::\@nil}
\def\mn@eprint@arXiv#1{\href {http://arxiv.org/abs/#1} {{\tt arXiv:#1}}}
\def\mn@eprint@dblp#1{\href {http://dblp.uni-trier.de/rec/bibtex/#1.xml}
  {dblp:#1}}
\def\mn@eprint@#1:#2:#3:#4\@nil{\def\@tempa {#1}\def\@tempb {#2}\def\@tempc
  {#3}\ifx \@tempc \@empty \let \@tempc \@tempb \let \@tempb \@tempa \fi \ifx
  \@tempb \@empty \def\@tempb {arXiv}\fi \@ifundefined
  {mn@eprint@\@tempb}{\@tempb:\@tempc}{\expandafter \expandafter \csname
  mn@eprint@\@tempb\endcsname \expandafter{\@tempc}}}

\bibitem[\protect\citeauthoryear{{Asplund}, {Grevesse}, {Sauval}  \&
  {Scott}}{{Asplund} et~al.}{2009}]{asplund2009}
{Asplund} M.,  {Grevesse} N.,  {Sauval} A.~J.,   {Scott} P.,  2009, \mn@doi
  [\araa] {10.1146/annurev.astro.46.060407.145222}, \href
  {https://ui.adsabs.harvard.edu/abs/2009ARA&A..47..481A} {47, 481}

\bibitem[\protect\citeauthoryear{{Aver}, {Olive}  \& {Skillman}}{{Aver}
  et~al.}{2010}]{AOS}
{Aver} E.,  {Olive} K.~A.,   {Skillman} E.~D.,  2010, \mn@doi [\jcap]
  {10.1088/1475-7516/2010/05/003}, \href
  {https://ui.adsabs.harvard.edu/abs/2010JCAP...05..003A} {2010, 003}

\bibitem[\protect\citeauthoryear{{Aver}, {Olive}  \& {Skillman}}{{Aver}
  et~al.}{2011}]{AOS2}
{Aver} E.,  {Olive} K.~A.,   {Skillman} E.~D.,  2011, \mn@doi [\jcap]
  {10.1088/1475-7516/2011/03/043}, \href
  {https://ui.adsabs.harvard.edu/abs/2011JCAP...03..043A} {2011, 043}

\bibitem[\protect\citeauthoryear{{Aver}, {Olive}  \& {Skillman}}{{Aver}
  et~al.}{2012}]{AOS3}
{Aver} E.,  {Olive} K.~A.,   {Skillman} E.~D.,  2012, \mn@doi [\jcap]
  {10.1088/1475-7516/2012/04/004}, \href
  {https://ui.adsabs.harvard.edu/abs/2012JCAP...04..004A} {2012, 004}

\bibitem[\protect\citeauthoryear{{Aver}, {Olive}, {Porter}  \&
  {Skillman}}{{Aver} et~al.}{2013}]{AOPS}
{Aver} E.,  {Olive} K.~A.,  {Porter} R.~L.,   {Skillman} E.~D.,  2013, \mn@doi
  [\jcap] {10.1088/1475-7516/2013/11/017}, \href
  {https://ui.adsabs.harvard.edu/abs/2013JCAP...11..017A} {2013, 017}

\bibitem[\protect\citeauthoryear{{Aver}, {Olive}  \& {Skillman}}{{Aver}
  et~al.}{2015}]{AOS4}
{Aver} E.,  {Olive} K.~A.,   {Skillman} E.~D.,  2015, \mn@doi [\jcap]
  {10.1088/1475-7516/2015/07/011}, \href
  {https://ui.adsabs.harvard.edu/abs/2015JCAP...07..011A} {2015, 011}

\bibitem[\protect\citeauthoryear{{Aver}, {Berg}, {Olive}, {Pogge}, {Salzer}  \&
  {Skillman}}{{Aver} et~al.}{2021}]{AOS5}
{Aver} E.,  {Berg} D.~A.,  {Olive} K.~A.,  {Pogge} R.~W.,  {Salzer} J.~J.,
  {Skillman} E.~D.,  2021, \mn@doi [\jcap] {10.1088/1475-7516/2021/03/027},
  \href {https://ui.adsabs.harvard.edu/abs/2021JCAP...03..027A} {2021, 027}

\bibitem[\protect\citeauthoryear{{Berg} et~al.,}{{Berg}
  et~al.}{2012}]{berg2012}
{Berg} D.~A.,  et~al., 2012, \mn@doi [\apj] {10.1088/0004-637X/754/2/98}, \href
  {https://ui.adsabs.harvard.edu/abs/2012ApJ...754...98B} {754, 98}

\bibitem[\protect\citeauthoryear{{Berg}, {Skillman}, {Croxall}, {Pogge},
  {Moustakas}  \& {Johnson-Groh}}{{Berg} et~al.}{2015}]{berg2015}
{Berg} D.~A.,  {Skillman} E.~D.,  {Croxall} K.~V.,  {Pogge} R.~W.,  {Moustakas}
  J.,   {Johnson-Groh} M.,  2015, \mn@doi [\apj] {10.1088/0004-637X/806/1/16},
  \href {https://ui.adsabs.harvard.edu/abs/2015ApJ...806...16B} {806, 16}

\bibitem[\protect\citeauthoryear{{Bohlin}}{{Bohlin}}{2010}]{bohlin10}
{Bohlin} R.~C.,  2010, \mn@doi [\aj] {10.1088/0004-6256/139/4/1515}, \href
  {https://ui.adsabs.harvard.edu/abs/2010AJ....139.1515B} {139, 1515}

\bibitem[\protect\citeauthoryear{{Cannon} et~al.,}{{Cannon}
  et~al.}{2011}]{cannon2011}
{Cannon} J.~M.,  et~al., 2011, \mn@doi [\apjl] {10.1088/2041-8205/739/1/L22},
  \href {https://ui.adsabs.harvard.edu/abs/2011ApJ...739L..22C} {739, L22}

\bibitem[\protect\citeauthoryear{{Cardelli}, {Clayton}  \& {Mathis}}{{Cardelli}
  et~al.}{1989}]{ccm89}
{Cardelli} J.~A.,  {Clayton} G.~C.,   {Mathis} J.~S.,  1989, \mn@doi [\apj]
  {10.1086/167900}, \href
  {https://ui.adsabs.harvard.edu/abs/1989ApJ...345..245C} {345, 245}

\bibitem[\protect\citeauthoryear{{Cyburt}, {Fields}  \& {Olive}}{{Cyburt}
  et~al.}{2002}]{cfo2}
{Cyburt} R.~H.,  {Fields} B.~D.,   {Olive} K.~A.,  2002, \mn@doi [Astroparticle
  Physics] {10.1016/S0927-6505(01)00171-2}, \href
  {https://ui.adsabs.harvard.edu/abs/2002APh....17...87C} {17, 87}

\bibitem[\protect\citeauthoryear{{Cyburt}, {Fields}, {Olive}  \&
  {Skillman}}{{Cyburt} et~al.}{2005}]{cfos}
{Cyburt} R.~H.,  {Fields} B.~D.,  {Olive} K.~A.,   {Skillman} E.,  2005,
  \mn@doi [Astroparticle Physics] {10.1016/j.astropartphys.2005.01.005}, \href
  {https://ui.adsabs.harvard.edu/abs/2005APh....23..313C} {23, 313}

\bibitem[\protect\citeauthoryear{{Cyburt}, {Fields}, {Olive}  \&
  {Yeh}}{{Cyburt} et~al.}{2016}]{cfoy}
{Cyburt} R.~H.,  {Fields} B.~D.,  {Olive} K.~A.,   {Yeh} T.-H.,  2016, \mn@doi
  [Reviews of Modern Physics] {10.1103/RevModPhys.88.015004}, \href
  {https://ui.adsabs.harvard.edu/abs/2016RvMP...88a5004C} {88, 015004}

\bibitem[\protect\citeauthoryear{{Davidson} \& {Kinman}}{{Davidson} \&
  {Kinman}}{1985}]{dk85}
{Davidson} K.,  {Kinman} T.~D.,  1985, \mn@doi [\apjs] {10.1086/191044}, \href
  {https://ui.adsabs.harvard.edu/abs/1985ApJS...58..321D} {58, 321}

\bibitem[\protect\citeauthoryear{{De Robertis}, {Dufour}  \& {Hunt}}{{De
  Robertis} et~al.}{1987}]{derob1987}
{De Robertis} M.~M.,  {Dufour} R.~J.,   {Hunt} R.~W.,  1987, \jrasc, \href
  {https://ui.adsabs.harvard.edu/abs/1987JRASC..81..195D} {81, 195}

\bibitem[\protect\citeauthoryear{{Dinerstein} \& {Shields}}{{Dinerstein} \&
  {Shields}}{1986}]{ds86}
{Dinerstein} H.~L.,  {Shields} G.~A.,  1986, \mn@doi [\apj] {10.1086/164753},
  \href {https://ui.adsabs.harvard.edu/abs/1986ApJ...311...45D} {311, 45}

\bibitem[\protect\citeauthoryear{{Ekta} \& {Chengalur}}{{Ekta} \&
  {Chengalur}}{2010}]{ekta2010}
{Ekta} B.,  {Chengalur} J.~N.,  2010, \mn@doi [\mnras]
  {10.1111/j.1365-2966.2010.16756.x}, \href
  {https://ui.adsabs.harvard.edu/abs/2010MNRAS.406.1238E} {406, 1238}

\bibitem[\protect\citeauthoryear{{Fern{\'a}ndez}, {Terlevich}, {D{\'\i}az}  \&
  {Terlevich}}{{Fern{\'a}ndez} et~al.}{2019}]{ftdt2}
{Fern{\'a}ndez} V.,  {Terlevich} E.,  {D{\'\i}az} A.~I.,   {Terlevich} R.,
  2019, \mn@doi [\mnras] {10.1093/mnras/stz1433}, \href
  {https://ui.adsabs.harvard.edu/abs/2019MNRAS.487.3221F} {487, 3221}

\bibitem[\protect\citeauthoryear{{Fields}, {Olive}, {Yeh}  \& {Young}}{{Fields}
  et~al.}{2020}]{fields2020}
{Fields} B.~D.,  {Olive} K.~A.,  {Yeh} T.-H.,   {Young} C.,  2020, \mn@doi
  [\jcap] {10.1088/1475-7516/2020/03/010}, \href
  {https://ui.adsabs.harvard.edu/abs/2020JCAP...03..010F} {2020, 010}

\bibitem[\protect\citeauthoryear{{Garnett}}{{Garnett}}{1992}]{garnett92}
{Garnett} D.~R.,  1992, \mn@doi [\aj] {10.1086/116146}, \href
  {https://ui.adsabs.harvard.edu/abs/1992AJ....103.1330G} {103, 1330}

\bibitem[\protect\citeauthoryear{{Giovanelli} et~al.,}{{Giovanelli}
  et~al.}{2005}]{giovanelli2005}
{Giovanelli} R.,  et~al., 2005, \mn@doi [\aj] {10.1086/497431}, \href
  {https://ui.adsabs.harvard.edu/abs/2005AJ....130.2598G} {130, 2598}

\bibitem[\protect\citeauthoryear{{Guseva}, {Izotov}, {Fricke}  \&
  {Henkel}}{{Guseva} et~al.}{2015}]{guseva2015}
{Guseva} N.~G.,  {Izotov} Y.~I.,  {Fricke} K.~J.,   {Henkel} C.,  2015, \mn@doi
  [\aap] {10.1051/0004-6361/201525697}, \href
  {https://ui.adsabs.harvard.edu/abs/2015A&A...579A..11G} {579, A11}

\bibitem[\protect\citeauthoryear{{Guseva}, {Izotov}, {Fricke}  \&
  {Henkel}}{{Guseva} et~al.}{2017}]{guseva2017}
{Guseva} N.~G.,  {Izotov} Y.~I.,  {Fricke} K.~J.,   {Henkel} C.,  2017, \mn@doi
  [\aap] {10.1051/0004-6361/201629181}, \href
  {https://ui.adsabs.harvard.edu/abs/2017A&A...599A..65G} {599, A65}

\bibitem[\protect\citeauthoryear{{Haynes} et~al.,}{{Haynes}
  et~al.}{2011}]{haynes2011}
{Haynes} M.~P.,  et~al., 2011, \mn@doi [\aj] {10.1088/0004-6256/142/5/170},
  \href {https://ui.adsabs.harvard.edu/abs/2011AJ....142..170H} {142, 170}

\bibitem[\protect\citeauthoryear{{Hirschauer} et~al.,}{{Hirschauer}
  et~al.}{2016}]{hirschauer2016}
{Hirschauer} A.~S.,  et~al., 2016, \mn@doi [\apj]
  {10.3847/0004-637X/822/2/108}, \href
  {https://ui.adsabs.harvard.edu/abs/2016ApJ...822..108H} {822, 108}

\bibitem[\protect\citeauthoryear{{Hsyu}, {Cooke}, {Prochaska}  \&
  {Bolte}}{{Hsyu} et~al.}{2018}]{hsyu2018}
{Hsyu} T.,  {Cooke} R.~J.,  {Prochaska} J.~X.,   {Bolte} M.,  2018, \mn@doi
  [\apj] {10.3847/1538-4357/aad18a}, \href
  {https://ui.adsabs.harvard.edu/abs/2018ApJ...863..134H} {863, 134}

\bibitem[\protect\citeauthoryear{{Hsyu}, {Cooke}, {Prochaska}  \&
  {Bolte}}{{Hsyu} et~al.}{2020}]{hcpb}
{Hsyu} T.,  {Cooke} R.~J.,  {Prochaska} J.~X.,   {Bolte} M.,  2020, \mn@doi
  [\apj] {10.3847/1538-4357/ab91af}, \href
  {https://ui.adsabs.harvard.edu/abs/2020ApJ...896...77H} {896, 77}

\bibitem[\protect\citeauthoryear{{Izotov} \& {Thuan}}{{Izotov} \&
  {Thuan}}{1999}]{izotov99}
{Izotov} Y.~I.,  {Thuan} T.~X.,  1999, \mn@doi [\apj] {10.1086/306708}, \href
  {https://ui.adsabs.harvard.edu/abs/1999ApJ...511..639I} {511, 639}

\bibitem[\protect\citeauthoryear{{Izotov} \& {Thuan}}{{Izotov} \&
  {Thuan}}{2007}]{izotov07b}
{Izotov} Y.~I.,  {Thuan} T.~X.,  2007, \mn@doi [\apj] {10.1086/519922}, \href
  {https://ui.adsabs.harvard.edu/abs/2007ApJ...665.1115I} {665, 1115}

\bibitem[\protect\citeauthoryear{{Izotov}, {Lipovetsky}, {Chaffee}, {Foltz},
  {Guseva}  \& {Kniazev}}{{Izotov} et~al.}{1997}]{izotov97}
{Izotov} Y.~I.,  {Lipovetsky} V.~A.,  {Chaffee} F.~H.,  {Foltz} C.~B.,
  {Guseva} N.~G.,   {Kniazev} A.~Y.,  1997, \mn@doi [\apj] {10.1086/303664},
  \href {https://ui.adsabs.harvard.edu/abs/1997ApJ...476..698I} {476, 698}

\bibitem[\protect\citeauthoryear{{Izotov}, {Thuan}  \& {Guseva}}{{Izotov}
  et~al.}{2005}]{izotov05}
{Izotov} Y.~I.,  {Thuan} T.~X.,   {Guseva} N.~G.,  2005, \mn@doi [\apj]
  {10.1086/432874}, \href
  {https://ui.adsabs.harvard.edu/abs/2005ApJ...632..210I} {632, 210}

\bibitem[\protect\citeauthoryear{{Izotov}, {Stasi{\'n}ska}, {Meynet}, {Guseva}
  \& {Thuan}}{{Izotov} et~al.}{2006}]{iz06}
{Izotov} Y.~I.,  {Stasi{\'n}ska} G.,  {Meynet} G.,  {Guseva} N.~G.,   {Thuan}
  T.~X.,  2006, \mn@doi [\aap] {10.1051/0004-6361:20053763}, \href
  {https://ui.adsabs.harvard.edu/abs/2006A&A...448..955I} {448, 955}

\bibitem[\protect\citeauthoryear{{Izotov}, {Thuan}  \&
  {Stasi{\'n}ska}}{{Izotov} et~al.}{2007}]{its07}
{Izotov} Y.~I.,  {Thuan} T.~X.,   {Stasi{\'n}ska} G.,  2007, \mn@doi [\apj]
  {10.1086/513601}, \href
  {https://ui.adsabs.harvard.edu/abs/2007ApJ...662...15I} {662, 15}

\bibitem[\protect\citeauthoryear{{Izotov}, {Thuan}  \& {Guseva}}{{Izotov}
  et~al.}{2012}]{itg12}
{Izotov} Y.~I.,  {Thuan} T.~X.,   {Guseva} N.~G.,  2012, \mn@doi [\aap]
  {10.1051/0004-6361/201219733}, \href
  {https://ui.adsabs.harvard.edu/abs/2012A&A...546A.122I} {546, A122}

\bibitem[\protect\citeauthoryear{{Izotov}, {Thuan}  \& {Guseva}}{{Izotov}
  et~al.}{2014}]{itg14}
{Izotov} Y.~I.,  {Thuan} T.~X.,   {Guseva} N.~G.,  2014, \mn@doi [\mnras]
  {10.1093/mnras/stu1771}, \href
  {https://ui.adsabs.harvard.edu/abs/2014MNRAS.445..778I} {445, 778}

\bibitem[\protect\citeauthoryear{{Izotov}, {Thuan}, {Guseva}  \&
  {Liss}}{{Izotov} et~al.}{2018a}]{izotov2018}
{Izotov} Y.~I.,  {Thuan} T.~X.,  {Guseva} N.~G.,   {Liss} S.~E.,  2018a,
  \mn@doi [\mnras] {10.1093/mnras/stx2478}, \href
  {https://ui.adsabs.harvard.edu/abs/2018MNRAS.473.1956I} {473, 1956}

\bibitem[\protect\citeauthoryear{{Izotov}, {Thuan}, {Guseva}  \&
  {Liss}}{{Izotov} et~al.}{2018b}]{itgl18}
{Izotov} Y.~I.,  {Thuan} T.~X.,  {Guseva} N.~G.,   {Liss} S.~E.,  2018b,
  \mn@doi [\mnras] {10.1093/mnras/stx2478}, \href
  {https://ui.adsabs.harvard.edu/abs/2018MNRAS.473.1956I} {473, 1956}

\bibitem[\protect\citeauthoryear{{James}, {Koposov}, {Stark}, {Belokurov},
  {Pettini}, {Olszewski}  \& {McQuinn}}{{James} et~al.}{2017}]{james2017}
{James} B.~L.,  {Koposov} S.~E.,  {Stark} D.~P.,  {Belokurov} V.,  {Pettini}
  M.,  {Olszewski} E.~W.,   {McQuinn} K. B.~W.,  2017, \mn@doi [\mnras]
  {10.1093/mnras/stw2962}, \href
  {https://ui.adsabs.harvard.edu/abs/2017MNRAS.465.3977J} {465, 3977}

\bibitem[\protect\citeauthoryear{{Kelson}}{{Kelson}}{2003}]{kelson03}
{Kelson} D.~D.,  2003, \mn@doi [\pasp] {10.1086/375502}, \href
  {https://ui.adsabs.harvard.edu/abs/2003PASP..115..688K} {115, 688}

\bibitem[\protect\citeauthoryear{{Kojima} et~al.,}{{Kojima}
  et~al.}{2020}]{kojima2020}
{Kojima} T.,  et~al., 2020, \mn@doi [\apj] {10.3847/1538-4357/aba047}, \href
  {https://ui.adsabs.harvard.edu/abs/2020ApJ...898..142K} {898, 142}

\bibitem[\protect\citeauthoryear{{Kurichin}, {Kislitsyn}, {Klimenko},
  {Balashev}  \& {Ivanchik}}{{Kurichin} et~al.}{2021}]{kurichin2021}
{Kurichin} O.~A.,  {Kislitsyn} P.~A.,  {Klimenko} V.~V.,  {Balashev} S.~A.,
  {Ivanchik} A.~V.,  2021, \mn@doi [\mnras] {10.1093/mnras/stab215}, \href
  {https://ui.adsabs.harvard.edu/abs/2021MNRAS.502.3045K} {502, 3045}

\bibitem[\protect\citeauthoryear{{Luridiana}, {Morisset}  \&
  {Shaw}}{{Luridiana} et~al.}{2012}]{luridiana12}
{Luridiana} V.,  {Morisset} C.,   {Shaw} R.~A.,  2012, \mn@doi [IAU Symposium]
  {10.1017/S1743921312011738}, \href
  {https://ui.adsabs.harvard.edu/abs/2012IAUS..283..422L} {283, 422}

\bibitem[\protect\citeauthoryear{{Luridiana}, {Morisset}  \&
  {Shaw}}{{Luridiana} et~al.}{2015}]{luridiana15}
{Luridiana} V.,  {Morisset} C.,   {Shaw} R.~A.,  2015, \mn@doi [\aap]
  {10.1051/0004-6361/201323152}, \href
  {https://ui.adsabs.harvard.edu/abs/2015A&A...573A..42L} {573, A42}

\bibitem[\protect\citeauthoryear{{Markov}}{{Markov}}{1906}]{mar}
{Markov} A.~A.,  1906, Proc. Phys. Math. Soc. Kazan Univ., 15, 135

\bibitem[\protect\citeauthoryear{{McQuinn} et~al.,}{{McQuinn}
  et~al.}{2015}]{mcquinn2015a}
{McQuinn} K. B.~W.,  et~al., 2015, \mn@doi [\apj]
  {10.1088/0004-637X/812/2/158}, \href
  {https://ui.adsabs.harvard.edu/abs/2015ApJ...812..158M} {812, 158}

\bibitem[\protect\citeauthoryear{{McQuinn} et~al.,}{{McQuinn}
  et~al.}{2020}]{mcquinn2020}
{McQuinn} K. B.~W.,  et~al., 2020, \mn@doi [\apj] {10.3847/1538-4357/ab7447},
  \href {https://ui.adsabs.harvard.edu/abs/2020ApJ...891..181M} {891, 181}

\bibitem[\protect\citeauthoryear{{Metropolis}, {Rosenbluth}, {Rosenbluth},
  {Teller}  \& {Teller}}{{Metropolis} et~al.}{1953}]{met}
{Metropolis} N.,  {Rosenbluth} A.~W.,  {Rosenbluth} M.~N.,  {Teller} A.~H.,
  {Teller} E.,  1953, \mn@doi [\jcp] {10.1063/1.1699114}, \href
  {https://ui.adsabs.harvard.edu/abs/1953JChPh..21.1087M} {21, 1087}

\bibitem[\protect\citeauthoryear{{Nava}, {Casebeer}, {Henry}  \&
  {Jevremovic}}{{Nava} et~al.}{2006}]{nava06}
{Nava} A.,  {Casebeer} D.,  {Henry} R. B.~C.,   {Jevremovic} D.,  2006, \mn@doi
  [\apj] {10.1086/504416}, \href
  {https://ui.adsabs.harvard.edu/abs/2006ApJ...645.1076N} {645, 1076}

\bibitem[\protect\citeauthoryear{{Oke}}{{Oke}}{1990}]{oke90}
{Oke} J.~B.,  1990, \mn@doi [\aj] {10.1086/115444}, \href
  {https://ui.adsabs.harvard.edu/abs/1990AJ.....99.1621O} {99, 1621}

\bibitem[\protect\citeauthoryear{{Olive} \& {Skillman}}{{Olive} \&
  {Skillman}}{2001}]{os01}
{Olive} K.~A.,  {Skillman} E.~D.,  2001, \mn@doi [\na]
  {10.1016/S1384-1076(01)00051-3}, \href
  {https://ui.adsabs.harvard.edu/abs/2001NewA....6..119O} {6, 119}

\bibitem[\protect\citeauthoryear{{Olive} \& {Skillman}}{{Olive} \&
  {Skillman}}{2004}]{os04}
{Olive} K.~A.,  {Skillman} E.~D.,  2004, \mn@doi [\apj] {10.1086/425170}, \href
  {https://ui.adsabs.harvard.edu/abs/2004ApJ...617...29O} {617, 29}

\bibitem[\protect\citeauthoryear{{Olive}, {Steigman}  \& {Walker}}{{Olive}
  et~al.}{2000}]{osw}
{Olive} K.~A.,  {Steigman} G.,   {Walker} T.~P.,  2000, \mn@doi [\physrep]
  {10.1016/S0370-1573(00)00031-4}, \href
  {https://ui.adsabs.harvard.edu/abs/2000PhR...333..389O} {333, 389}

\bibitem[\protect\citeauthoryear{{Pagel}, {Simonson}, {Terlevich}  \&
  {Edmunds}}{{Pagel} et~al.}{1992}]{pagel92}
{Pagel} B.~E.~J.,  {Simonson} E.~A.,  {Terlevich} R.~J.,   {Edmunds} M.~G.,
  1992, \mn@doi [\mnras] {10.1093/mnras/255.2.325}, \href
  {https://ui.adsabs.harvard.edu/abs/1992MNRAS.255..325P} {255, 325}

\bibitem[\protect\citeauthoryear{{Particle Data Group} et~al.,}{{Particle Data
  Group} et~al.}{2020}]{rpp}
{Particle Data Group} et~al., 2020, \mn@doi [Progress of Theoretical and
  Experimental Physics] {10.1093/ptep/ptaa104}, \href
  {https://ui.adsabs.harvard.edu/abs/2020PTEP.2020h3C01P} {2020, 083C01}

\bibitem[\protect\citeauthoryear{{Peimbert} \& {Costero}}{{Peimbert} \&
  {Costero}}{1969}]{peimbert69}
{Peimbert} M.,  {Costero} R.,  1969, Boletin de los Observatorios Tonantzintla
  y Tacubaya, \href {https://ui.adsabs.harvard.edu/abs/1969BOTT....5....3P} {5,
  3}

\bibitem[\protect\citeauthoryear{{Peimbert} \& {Torres-Peimbert}}{{Peimbert} \&
  {Torres-Peimbert}}{1974}]{ptp74}
{Peimbert} M.,  {Torres-Peimbert} S.,  1974, \mn@doi [\apj] {10.1086/153166},
  \href {https://ui.adsabs.harvard.edu/abs/1974ApJ...193..327P} {193, 327}

\bibitem[\protect\citeauthoryear{{Peimbert}, {Luridiana}  \&
  {Peimbert}}{{Peimbert} et~al.}{2007}]{plp07}
{Peimbert} M.,  {Luridiana} V.,   {Peimbert} A.,  2007, \mn@doi [\apj]
  {10.1086/520571}, \href
  {https://ui.adsabs.harvard.edu/abs/2007ApJ...666..636P} {666, 636}

\bibitem[\protect\citeauthoryear{{Pisanti}, {Cirillo}, {Esposito}, {Iocco},
  {Mangano}, {Miele}  \& {Serpico}}{{Pisanti} et~al.}{2008}]{pis}
{Pisanti} O.,  {Cirillo} A.,  {Esposito} S.,  {Iocco} F.,  {Mangano} G.,
  {Miele} G.,   {Serpico} P.~D.,  2008, \mn@doi [Computer Physics
  Communications] {10.1016/j.cpc.2008.02.015}, \href
  {https://ui.adsabs.harvard.edu/abs/2008CoPhC.178..956P} {178, 956}

\bibitem[\protect\citeauthoryear{{Pitrou}, {Coc}, {Uzan}  \&
  {Vangioni}}{{Pitrou} et~al.}{2018}]{coc18}
{Pitrou} C.,  {Coc} A.,  {Uzan} J.-P.,   {Vangioni} E.,  2018, \mn@doi
  [\physrep] {10.1016/j.physrep.2018.04.005}, \href
  {https://ui.adsabs.harvard.edu/abs/2018PhR...754....1P} {754, 1}

\bibitem[\protect\citeauthoryear{{Planck Collaboration} et~al.,}{{Planck
  Collaboration} et~al.}{2020}]{planck18}
{Planck Collaboration} et~al., 2020, \mn@doi [\aap]
  {10.1051/0004-6361/201833910}, \href
  {https://ui.adsabs.harvard.edu/abs/2020A&A...641A...6P} {641, A6}

\bibitem[\protect\citeauthoryear{{Pogge} et~al.,}{{Pogge}
  et~al.}{2010}]{pogge2010}
{Pogge} R.~W.,  et~al., 2010, in {McLean} I.~S.,  {Ramsay} S.~K.,   {Takami}
  H.,  eds,  Society of Photo-Optical Instrumentation Engineers (SPIE)
  Conference Series Vol. 7735, Ground-based and Airborne Instrumentation for
  Astronomy III. p. 77350A, \mn@doi{10.1117/12.857215}

\bibitem[\protect\citeauthoryear{{Pustilnik}, {Kniazev}  \&
  {Pramskij}}{{Pustilnik} et~al.}{2005}]{pustilnik05}
{Pustilnik} S.~A.,  {Kniazev} A.~Y.,   {Pramskij} A.~G.,  2005, \mn@doi [\aap]
  {10.1051/0004-6361:20053102}, \href
  {https://ui.adsabs.harvard.edu/abs/2005A&A...443...91P} {443, 91}

\bibitem[\protect\citeauthoryear{{Pustilnik}, {Kniazev}, {Perepelitsyna}  \&
  {Egorova}}{{Pustilnik} et~al.}{2020}]{pustilnik2020}
{Pustilnik} S.~A.,  {Kniazev} A.~Y.,  {Perepelitsyna} Y.~A.,   {Egorova} E.~S.,
   2020, \mn@doi [\mnras] {10.1093/mnras/staa215}, \href
  {https://ui.adsabs.harvard.edu/abs/2020MNRAS.493..830P} {493, 830}

\bibitem[\protect\citeauthoryear{{Rogers}, {Skillman}, {Pogge}, {Berg},
  {Moustakas}, {Croxall}  \& {Sun}}{{Rogers} et~al.}{2021}]{rogers2021}
{Rogers} N. S.~J.,  {Skillman} E.~D.,  {Pogge} R.~W.,  {Berg} D.~A.,
  {Moustakas} J.,  {Croxall} K.~V.,   {Sun} J.,  2021, \mn@doi [\apj]
  {10.3847/1538-4357/abf8b9}, \href
  {https://ui.adsabs.harvard.edu/abs/2021ApJ...915...21R} {915, 21}

\bibitem[\protect\citeauthoryear{{S{\'a}nchez Almeida}, {Filho}, {Dalla
  Vecchia}  \& {Skillman}}{{S{\'a}nchez Almeida} et~al.}{2017}]{sanchez2017}
{S{\'a}nchez Almeida} J.,  {Filho} M.~E.,  {Dalla Vecchia} C.,   {Skillman}
  E.~D.,  2017, \mn@doi [\apj] {10.3847/1538-4357/835/2/159}, \href
  {https://ui.adsabs.harvard.edu/abs/2017ApJ...835..159S} {835, 159}

\bibitem[\protect\citeauthoryear{{Schlafly} \& {Finkbeiner}}{{Schlafly} \&
  {Finkbeiner}}{2011}]{sf2011}
{Schlafly} E.~F.,  {Finkbeiner} D.~P.,  2011, \mn@doi [\apj]
  {10.1088/0004-637X/737/2/103}, \href
  {https://ui.adsabs.harvard.edu/abs/2011ApJ...737..103S} {737, 103}

\bibitem[\protect\citeauthoryear{{Seifert} et~al.,}{{Seifert}
  et~al.}{2003}]{seifert2003}
{Seifert} W.,  et~al., 2003, in {Iye} M.,  {Moorwood} A. F.~M.,  eds,  Society
  of Photo-Optical Instrumentation Engineers (SPIE) Conference Series Vol.
  4841, Instrument Design and Performance for Optical/Infrared Ground-based
  Telescopes. pp 962--973, \mn@doi{10.1117/12.459494}

\bibitem[\protect\citeauthoryear{{Senchyna} \& {Stark}}{{Senchyna} \&
  {Stark}}{2019}]{senchyna2019}
{Senchyna} P.,  {Stark} D.~P.,  2019, \mn@doi [\mnras] {10.1093/mnras/stz058},
  \href {https://ui.adsabs.harvard.edu/abs/2019MNRAS.484.1270S} {484, 1270}

\bibitem[\protect\citeauthoryear{{Senchyna}, {Stark}, {Chevallard}, {Charlot},
  {Jones}  \& {Vidal-Garc{\'\i}a}}{{Senchyna} et~al.}{2019}]{senchyna2019b}
{Senchyna} P.,  {Stark} D.~P.,  {Chevallard} J.,  {Charlot} S.,  {Jones} T.,
  {Vidal-Garc{\'\i}a} A.,  2019, \mn@doi [\mnras] {10.1093/mnras/stz1907},
  \href {https://ui.adsabs.harvard.edu/abs/2019MNRAS.488.3492S} {488, 3492}

\bibitem[\protect\citeauthoryear{{Skillman} \& {Kennicutt}}{{Skillman} \&
  {Kennicutt}}{1993}]{skillman93}
{Skillman} E.~D.,  {Kennicutt} Robert~C. J.,  1993, \mn@doi [\apj]
  {10.1086/172868}, \href
  {https://ui.adsabs.harvard.edu/abs/1993ApJ...411..655S} {411, 655}

\bibitem[\protect\citeauthoryear{{Skillman} et~al.,}{{Skillman}
  et~al.}{2013}]{skillman2013}
{Skillman} E.~D.,  et~al., 2013, \mn@doi [\aj] {10.1088/0004-6256/146/1/3},
  \href {https://ui.adsabs.harvard.edu/abs/2013AJ....146....3S} {146, 3}

\bibitem[\protect\citeauthoryear{{Stasi{\'n}ska}}{{Stasi{\'n}ska}}{1990}]{stasinska90}
{Stasi{\'n}ska} G.,  1990, \aaps, \href
  {https://ui.adsabs.harvard.edu/abs/1990A&AS...83..501S} {83, 501}

\bibitem[\protect\citeauthoryear{{Storey} \& {Hummer}}{{Storey} \&
  {Hummer}}{1995}]{sh95}
{Storey} P.~J.,  {Hummer} D.~G.,  1995, \mn@doi [\mnras]
  {10.1093/mnras/272.1.41}, \href
  {https://ui.adsabs.harvard.edu/abs/1995MNRAS.272...41S} {272, 41}

\bibitem[\protect\citeauthoryear{{Thuan}, {Izotov}  \& {Lipovetsky}}{{Thuan}
  et~al.}{1995}]{thuan95}
{Thuan} T.~X.,  {Izotov} Y.~I.,   {Lipovetsky} V.~A.,  1995, \mn@doi [\apj]
  {10.1086/175676}, \href
  {https://ui.adsabs.harvard.edu/abs/1995ApJ...445..108T} {445, 108}

\bibitem[\protect\citeauthoryear{{Tremonti} et~al.,}{{Tremonti}
  et~al.}{2004}]{tremonti2004}
{Tremonti} C.~A.,  et~al., 2004, \mn@doi [\apj] {10.1086/423264}, \href
  {https://ui.adsabs.harvard.edu/abs/2004ApJ...613..898T} {613, 898}

\bibitem[\protect\citeauthoryear{{Valerdi}, {Peimbert}, {Peimbert}  \&
  {Sixtos}}{{Valerdi} et~al.}{2019}]{vpps}
{Valerdi} M.,  {Peimbert} A.,  {Peimbert} M.,   {Sixtos} A.,  2019, \mn@doi
  [\apj] {10.3847/1538-4357/ab14e4}, \href
  {https://ui.adsabs.harvard.edu/abs/2019ApJ...876...98V} {876, 98}

\bibitem[\protect\citeauthoryear{{Walker}, {Steigman}, {Schramm}, {Olive}  \&
  {Kang}}{{Walker} et~al.}{1991}]{wssok}
{Walker} T.~P.,  {Steigman} G.,  {Schramm} D.~N.,  {Olive} K.~A.,   {Kang}
  H.-S.,  1991, \mn@doi [\apj] {10.1086/170255}, \href
  {https://ui.adsabs.harvard.edu/abs/1991ApJ...376...51W} {376, 51}

\bibitem[\protect\citeauthoryear{{Yang}, {Malhotra}, {Rhoads}  \&
  {Wang}}{{Yang} et~al.}{2017}]{yang2017}
{Yang} H.,  {Malhotra} S.,  {Rhoads} J.~E.,   {Wang} J.,  2017, \mn@doi [\apj]
  {10.3847/1538-4357/aa8809}, \href
  {https://ui.adsabs.harvard.edu/abs/2017ApJ...847...38Y} {847, 38}

\bibitem[\protect\citeauthoryear{Yeh, Olive  \& Fields}{Yeh
  et~al.}{2021}]{Yeh:2020mgl}
Yeh T.-H.,  Olive K.~A.,   Fields B.~D.,  2021, \mn@doi [JCAP]
  {10.1088/1475-7516/2021/03/046}, 03, 046

\bibitem[\protect\citeauthoryear{{van Zee} \& {Haynes}}{{van Zee} \&
  {Haynes}}{2006}]{vanzee2006}
{van Zee} L.,  {Haynes} M.~P.,  2006, \mn@doi [\apj] {10.1086/498017}, \href
  {https://ui.adsabs.harvard.edu/abs/2006ApJ...636..214V} {636, 214}

\makeatother
\end{thebibliography}

\label{lastpage}
\end{document}